\documentclass[11pt,dvipsnames]{article}

\usepackage{amsmath,amsfonts,amsthm,amssymb,color,bbm}
\usepackage{fancybox}
\usepackage{algorithm}
\usepackage[noend]{algpseudocode}
\usepackage{dsfont}
\usepackage{bm}
\usepackage{hyperref,cleveref}
\usepackage{subfig}
\usepackage[margin=1in]{geometry}
\usepackage{multirow}
\usepackage{thmtools}
\usepackage{thm-restate}
\usepackage{pgf,tikz}
\usepackage{graphicx}
\usepackage{etoolbox}
\usepackage{times}
\usepackage[makeroom]{cancel}

\usetikzlibrary{math,calc,arrows, shapes,positioning,decorations.pathreplacing}
\usepackage{tkz-graph}
\tikzset{
  LabelStyle/.style = { rectangle, rounded corners, draw,
                        minimum width = 2em, fill = yellow!50,
                        text = red, font = \bfseries },
  VertexStyle/.append style = { font = \bfseries},
  EdgeStyle/.append style = {->, bend right} 
  }
\tikzstyle{vertex}=[fill=black, draw=black, shape=circle, minimum size = 5pt,inner sep=0pt]

\newtheorem{theorem}{Theorem}
\newtheorem{lemma}[theorem]{Lemma}

\newtheorem{observation}[theorem]{Observation}

\newtheorem{definition}[theorem]{Definition}

  {\begin{Sbox}\begin{minipage}}%
  {\end{minipage}\end{Sbox}\fbox{\TheSbox}}

\newcommand{\set}[1]{\left\{#1\right\}}

\def\expec#1#2{{\mathbb{E}}_{#1}\left[ #2 \right]}
\newcommand{\E}{\mathbb{E}}
\def\defeq{\stackrel{\mathrm{def}}{=}}

\def\R{\mathbb{R}}

\newcommand{\beq}{\begin{equation}}
\newcommand{\eeq}{\end{equation}}

\usepackage{cite,etoolbox}
\newcommand\Alg{\text{ALG}}
\newcommand\Opt{\text{OPT}}
\renewcommand{\chi}{x}
\renewcommand{\psi}{y}
\renewcommand{\zeta}{z}

\newcommand{\Z}{\mathbb{Z}}

\makeatletter 
\pretocmd\@bibitem{\color{black}\csname keycolor#1\endcsname}{}{\fail}
\newcommand\citecolor[1]{\@namedef{keycolor#1}{\color{blue}}}
\makeatother
\citecolor{ChekuriM97}


\begin{document}

\title{
Multi Purpose Optimal Routing: New Perspectives and Algorithms 
}
\date{}

\author{
Majid Farhadi\thanks{Georgia Institute of Technology.
\texttt{farhadi@gatech.edu}.
}
\and
Jai Moondra\thanks{Georgia Institute of Technology.
\texttt{jmoondra3@gatech.edu}.
}
\and
Prasad Tetali\thanks{Carnegie Mellon University.
\texttt{ptetali@cmu.edu}.
}
\and
Alejandro Toriello\thanks{Georgia Institute of Technology.
\texttt{atoriello@isye.gatech.edu}.
}
}

\maketitle

\begin{abstract}
    The cost due to delay in services may be intrinsically different for various applications of vehicle routing such as medical emergencies, logistical operations, and ride-sharing. We study a fundamental generalization of the Traveling Salesman Problem, namely $L_p$~TSP, where the objective is to minimize an aggregated measure of the delay in services, quantified by the Minkowski $p$-norm of the delay vector. We present efficient combinatorial and Linear Programming algorithms for approximating $L_p$~TSP on general metrics. We provide several approximation algorithms for the $L_p$ TSP problem, including $4.27$ \textit{\&} $10.92$-approximation algorithms for single \text{\&} multi vehicle $L_2$~TSP, called the Traveling Firefighter Problem. Among other contributions, we provide an $8$-approximation and a $1.78$ inapproximability for All-Norm TSP problem, addressing scenarios where one does not know the ideal cost function, or is seeking simultaneous approximation with respect to any cost function. 
\end{abstract}

\section{Introduction}

The Traveling Salesman Problem (TSP) is a fundamental problem in combinatorial optimization, and its variants have numerous applications. A multitude of studies on this problem have led to new computational machinery. The common goal of these problems is to find an optimal permutation to visit a set of destinations. Applications ranging from ride sharing to microchip design and genome sequencing suggest different measures as the appropriate objective for the optimization problem.

For example, in devising the drop off route of a school bus, one can optimize the fuel consumption (i.e., the driver's time en route) or the students' average/total waiting time. More generally, different measures of fairness can differently penalize larger waiting times. As another example, in fighting against wildfires (under an abstract model adaptable for real-world applications), the cost due to delay in service is better modelled quadratically versus linearly, therefore minimizing the Euclidean norm of the vector of visit times would be a better objective. Likewise, this line of research can lead to applications of routing theory in socioeconomic and environmental contexts.

Incorporating the above examples, in this paper we study the $L_p$ Traveling Salesman Problem ($L_p$~TSP). The input consists of an origin, a set of destinations, and underlying distances. The output is a sequence of destinations, to be visited by a traveler of unit speed, starting from a specified origin. The objective is to minimize the Minkowski $p$-norm of the resulting vector of visit (service) times. Since the traveler has unit speed, we interchangeably use the terms ``time'' and ``distance''. We also consider the variant of the problem where multiple vehicles are available, dispatched from specified locations, and dub it the \emph{multi-vehicle} $L_p$ TSP.

For $p = \infty$ the problem is a path variant of the celebrated Traveling Salesman Problem, and for $p = 1$ it is the classical Traveling Repairman Problem (TRP) \cite{ACPPP86}. For $p = 2$, the problem asks to minimize the Euclidean norm of the visit times vector, i.e., penalizing delays quadratically, and we call it the `Traveling Firefighter Problem' (TFP). As a generalization of these problems, $L_p$~TSP introduces a continuous spectrum of problems with many potential applications, in addition to raising new theoretical questions on relations, similarities, and differences between these optimization problems. 

The set of feasible routes for all $L_p$~TSP problems is the same, but an optimal solution for one is not necessarily optimal for another. Even the computational complexity of the routing problem is affected by the norm. For instance, TRP is strongly NP-hard for tree metrics \cite{S02}, i.e., when the distance function is defined by shortest paths along edges of a tree over the destinations. In contrast, finding TSP of tree metrics is trivially in P. For general metrics, there is a constant $\gamma > 1$ such that $L_p$~TSP problems are hard to $\gamma$-approximate (see \cite{KLS15} and references therein). The theory of computing and approximation algorithms has dedicated enormous efforts to understand the computational complexity of extremal cases of $L_p$~TSP, i.e., $L_1$~TSP \cite{BCCPRS94,GK98, CGRT03, GGKT08,bienkowski2021traveling} and $L_{\infty}$~TSP \cite{Wol80, SW90, BP91, Goe95, CV00, GLS05, BC11, SWZ12, HNR19,KKO20}.

We develop multiple techniques to tackle $L_p$~TSP  and related problems. Here we overview an outlook on the objectives of this study and mention some key results, to be formally presented in Section~\ref{sec:formal-results}.

\noindent\textbf{Online / All-Norm TSP.} In many situations, we may not know the right objective in advance, or there may be a Pareto efficiency trade-off. An interesting objective can be to search for a route which is competitive with respect to a norm of the delay vector that is determined by an adversary. That is, the route should be approximately optimal with respect to any norm of the delay vector. Similar concept of all norm optimization has been of interest in other scenarios, e.g., in routing, load balancing \cite{KRT01}, and machine scheduling \cite{AERW04,BP03}.
We develop an $8+ \varepsilon$-approximation algorithm for All-Norm TSP in general metric spaces, improving upon the previous $16$-approximation by \cite{GGKT08}.
We further show that an approximate All-Norm TSP cannot be necessarily guaranteed to within a factor less than $1.78$, that is the first of its kind for this problem, to the best of our knowledge.

\noindent\textbf{Segmented TSP.}
Different objectives challenge adaptability of the algorithms designed for one to be used for another. Using dynamic programming, we develop an approximation preserving reduction (that is, adding a multiplicative error of $1+\varepsilon$) from any $L_p$~TSP to (polynomially many instances of) a unique routing problem, known as Segmented TSP, introduced by Sitters \cite{S14}. The input to Segmented TSP includes a number of deadlines by which given numbers of vertices should be visited. The number of deadlines varies only with $p$ and $O(\varepsilon^{-2})$ and is independent of the underlying metric. Our theory along with the results from \cite{S14} enable $1+\varepsilon$ Polynomial Time Approximation Scheme (PTAS) for $L_p$~TSP on tree and Euclidean metrics and further motivate study of approximation algorithms for Segmented~TSP. 

\noindent\textbf{Constant Factor Approximation.} An $(8+\varepsilon)$-approximation of any $L_p$~TSP is immediate due to our $(8 + \varepsilon)$-approximation of All-Norm TSP. We further show that the ``kernel'' of our combinatorial algorithm can be reverse engineered with respect to performance guarantees for any specific norm. Analytically solving the consequent min max optimization problem, we provide a combinatorial $4.27$-approximation algorithm for the Traveling Firefighter Problem, i.e., $L_2$ TSP, and 
$\frac{8}{(p \ln 4)^{1/p}}$-approximation for $L_p$~TSP.

\noindent\textbf{LP Rounding And Multi-Vehicle Routing.}
A natural generalization of $L_p$~TSP, that is essential in many applications \cite{jmal2019apply,kulichmulti20}, is when we have multiple vehicles. For example, dispatching school buses, the total time spent in the traffic is a function of the assignment of students to buses as well as the bus routes. Likewise, firefighters can be partitioned into multiple teams and dispatched along different routes. The combinatorial algorithms fall short of generalizing to multi-vehicle scenario. We extend the theory of approximability to this case, using Linear Programming (LP) relaxations for multi-vehicle $L_p$~TSP. Our LP-based method also matches the approximation bounds for the combinatorial algorithm for the single vehicle problem. This further improves the scalability of the algorithms for single vehicle case by enabling use of advanced optimization machinery. We provide first constant factor approximations (parameterized by $p$) for multi vehicle $L_p$ TSP including a $10.92$-approximation for the multi Traveling Firefighters Problem. 
\footnote{A preliminary version of this article appeared in the Proceedings of SIAM Conference on Applied and Computational Discrete Algorithms (ACDA21) \cite{FTT21}.}

\subsection{On TSP \& TRP Literature}

Traveling Salesman Problem (TSP) is a principal problem in computer science, combinatorial optimization, and operations research, and its first formulations date back to as early as $19$\textsuperscript{th} century (c.f.\ \cite{ABCC06}).
The objective for this most famous routing problem is to visit a set of destinations and to return to the origin as soon as possible. Denoting the set (of destinations and the origin) by $V = [n]$ along with a distance function $d: V \times V \rightarrow \mathbb{R}_{\ge 0}$, TSP can be formulated as $\min_{\sigma \in [n]!} d(\sigma_n, \sigma_1) + \sum_{i \in [n-1]} d(\sigma_i, \sigma_{i+1})$, where $\sigma$ denotes the order in which the destinations are visited. Finding a solution within any constant factor of the optimal objective for TSP in polynomial time is not possible, unless P = NP. The problem becomes more tractable if we assume that $d$ is a metric, i.e., the distances satisfy triangle inequality, i.e., $d(i,j)+d(j,k) \ge d(i,k) \quad \forall i,j,k \in [n]$. This is equivalent to allowing the vehicle to revisit vertices, or having $d$ as the length of the shortest path between pairs of vertices along a weighted graph with vertex set $V$. The metric assumption also allows matching approximation bounds for when the traveller is not required to return to the origin, i.e., when the problem is to find strolls or paths instead of tours \cite{TVZ20}. Since the celebrated $3/2$-approximation algorithm of Christofides-Serdyukov \cite{Chr76,Ser78}, for its tour-variant on general metrics, TSP has been extensively studied for half a century \cite{Wol80, SW90, BP91, Goe95, CV00, GLS05, BC11, SWZ12, HNR19}. Very recently, Karlin, Klein, and Oveis Gharan \cite{KKO20} showed TSP can be approximated strictly better than $3/2$, while the problem is NP-hard to approximate within a factor of $123/122$ \cite{KLS15}.

TSP can fall short of modeling many real-world routing applications, and studying variants and generalizations of this fundamental problem is an active area of research. For instance, the traveling purchaser problem \cite{xiao2020traveling} requires determining a tour of the graph to purchase commodities from different supplier nodes. Frequently, the objective is modified to incorporate other constraints on the tour, such as the Steiner traveling salesman problem \cite{zhang2015steinertsp} and the prize-collecting TSP \cite{hubert2020prizecollectingtsp}. Routing questions have also been studied in continuous metrics (as opposed to discrete, graph-induced metrics) \cite{luo2019datacollection}. Some online routing problems require `servicing' vertices as requests arrive at different times, with penalties for any delay in service \cite{azar2021osd, ausiello2008onlinetsp}.

The common ground in variants of TSP is that the destinations are to be visited as fast as possible, i.e., optimizing the time spent by the server/traveler. In contrast, a classical counterpart to TSP is the Traveling Repairman Problem (TRP), a.k.a. the Minimum Latency Problem, the school bus driver problem \cite{will1994extremal}, hidden treasure \cite{BCCPRS94,koutsoupias1996searching,ausiello2000salesmen}, and the deliveryman problem \cite{minieka1989delivery,fischetti1993delivery,mendez2008new}. This problem requires optimizing the route from the perspective of clients, i.e., the total waiting time to be visited, and is another extensively studied combinatorial optimization problem \cite{ACPPP86, papadimitriou1993traveling, BCCPRS94, GK98, CGRT03, AW03} with a state-of-the-art approximation factor of $\simeq 3.59$ for general metrics \cite{CGRT03,PS14}. TRP requires the average visit distance for the vertices to be minimized. TRP has turned out to be even harder than TSP in certain aspects. For example, it is strongly NP-hard for (metrics defined by shortest paths along edges of) tree metrics \cite{S02}, where TSP is linear time solvable. For more literature on the TRP, refer to \cite{S21} and references therein.

The cost function for TSP can be written as $\max_{v \in V} \ell_v^{\sigma}$, where $\ell^{\sigma} \in \mathbb{R}^{V}$ is the vector of visit times of the vertices. That is, TSP seeks to minimize the $L_\infty$-norm of the vector of visit times, just as TRP seeks to minimize the $L_1$-norm of this vector. The ideal cost function over the visit times may differ for various applications. We suggest quadratic penalization of delays in containment of wildfires, and name the corresponding problem to minimize $\|\ell^{\sigma}\|_2$ as the Traveling Firefighter Problem (TFP). More generally, for $p \ge 1$, the $L_p$~TSP problem seeks to minimize the $L_p$-norm of the visit times. Indeed, containment of fires can be abstracted from various perspectives. Hartnell \cite{hartnell1995firefighter} modeled a constant speed spread of fires through edges of a graph, along which the firefighters also displace. Many objectives, such as minimization of the number of burned vertices, or the required time to contain the fire(s) are studied in this model and the problem is an active area of research. See \cite{finbow2009firefighter, klein2014approximation,adjiashvili2018firefighting,amir2020firefighter,deutch2021multi} and references therein. 

Multi-objective combinatorial optimization, and in particular multi-norm optimization has received significant attention because of its usefulness in modelling disparate requirements, fair policy making, and theoretical interest \cite{chakrabarty2018ordered-k-median, ibrahimpur2020stochastic, ehrgott2003multiobjective, KRT01, AERW04, BP03}. 

Generalizing the objective to Minkowski norm of the solution has allowed interpolating other classical problems in combinatorial optimization, e.g., $L_p$ set cover problem \cite{GGKT08,BBFT21} that further united the greedy algorithms for set-cover and the minimum-sum-set-cover \cite{FLT04} problems. Set cover is better approximable for $p = 1$ than $p = \infty\,,$ while TSP ($p = \infty$) is currently better approximated than TRP ($p = 1$). Nevertheless, this order is not expected to be reversed as TRP is an intrinsically harder problem. Moreover, in contrast to the concordance among $L_p$ set cover problems, for which the same greedy algorithm gives best (possible) bounds for any $p \in [1, \infty)$, state-of-the-art algorithms for TSP and TRP are noticeably different, and both are potentially far away from best possible approximation algorithms. This is yet another motivation to study $L_p$~TSP, ultimately towards unified best algorithms for routing problems.

\subsection{Problem Definition \& Examples}

Adding to the notation, let $[z]$ denote the set of numbers $\set{1, \cdots, z}$ for any positive integer $z$. $\tilde{O}(\cdot)$ is equivalent to $O(\cdot)$, treating $\varepsilon > 0$ as a constant. Recall that a metric over a set of nodes $V$ is a distance function $d: V \times V \rightarrow \mathbb{R}_{\ge 0}$ that satisfies symmetry, $d(x,y) = d(y,x) \quad \forall x, y \in V$, identity, $d(x,x) = 0$, and the triangle inequality,
$
d(x,y) \leq d(x,z) + d(z,y) \quad \forall x,y,z \in V\,.
$

The inputs to single-vehicle problems are a set of vertices $V$, including the destinations and the traveler's starting location, $s \in V$, and the underlying metric $d(\cdot,\cdot)$ over $V$, corresponding to distances (or the time it takes to travel) between pairs of vertices. 

The output will be from a set of feasible solutions/routes, denoted $\mathcal{F}$, that is permutations of $V$ that have the origin as their first element. Denoting the $i$\textsuperscript{th} visited vertex by $\sigma_i$, note that $\sigma_1 = s$. $T^\sigma_i$ is the $i$\textsuperscript{th} smallest visit time, due to a feasible solution $\sigma \in \mathcal{F}$, which can be written as $T^\sigma_1 = 0$ and $T^\sigma_i = \sum_{j = 2}^{i} d(\sigma_{j-1},\sigma_j)$ for $i \in \{2,\cdots,n\}$. The visit time for vertex $v$ is denoted $\ell^\sigma_v$. Similarly, $\ell^\sigma_s = 0$ and $\ell^\sigma_v = \sum_{i=2}^{\sigma_i = v} d(\sigma_{i-1},\sigma_i)  \quad \forall v \ne s$.

\begin{definition}[$L_p$~TSP]
The input of the optimization problem $L_p$~TSP is a set of destinations $V$, a starting vertex $s \in V$, and a metric $d: V \times V \rightarrow \mathbb{R}_{\ge 0}$. 
The objective is to find a feasible route $\sigma \in \mathcal{F}$, i.e., starting at $s$ and visiting all $v \in V$, that minimizes the Minkowski $p$-norm of the visit times, i.e.\ $\min_{\sigma \in \mathcal{F}} \|\ell^\sigma\|_p$, where
$
\|\ell^\sigma\|_p := \left(\sum_{v \in V} |\ell^\sigma_v|^p\right)^{\frac{1}{p}}\,.
$
\end{definition}

For the multi-vehicle variant, given $K$ vehicles starting at vertices $s_1, \ldots, s_K$ respectively, the multi-vehicle $L_p$~TSP seeks to minimize the $L_p$-norm of the vector of visit times of the destinations by any of the vehicles. We deal with multi-vehicle $L_p$~TSP only in Section \ref{sec: lp}, and unless otherwise stated, we assume to have a single vehicle.
When the problem objective is clear from context, we denote an optimal route as $\Opt$ and the answer of our algorithm as $\Alg$.

\noindent\textbf{A Spectrum of Candidate Problems.}
$L_p$~TSP enables a smooth transition between two extreme problems/objectives. For larger values of $p$ the objective is strongly affected by the dominating (larger) entries of the delay vector and minimizing $\|\ell\|_p$ is more efficient from the perspective of the server. In contrast, smaller values of $p$ provide a further uniformly aggregated cost due to the amount of time that the customers have waited for the service. This trade-off can also be interpreted from a fairness perspective, as increasing $p$ discourages the longest waiting time from becoming too large.

\noindent\textbf{Firefighter Example.}
We further elaborate on why $p \notin \set{1, \infty}$ can be a useful objective by considering the routing of a firefighter.\footnote{Over the past decade, and for the first time on record, the annual number of acres burned in the United States exceeded $10$ million; and it occurred twice \cite{Hoover18}. During $2018$, wildfire damages only in California totalled $\$150$ billion \cite{wang2021economic}. In July $2019$, a record $2.4$ million acres of the Amazon rainforest were torched \cite{natgeo}. The optimal allocation, scheduling, and routing of firefighting resources may help to better address this global challenge.} Consider a set of wildfires in dispersed locations, and suppose a skilled firefighter extinguishes any fire the second they arrive at a location; the firefighter must choose the order in which to visit and extinguish the fires. One possible strategy is to choose a sequence in order to finish extinguishing all fires as soon as possible, which corresponds to solving the $L_\infty$~TSP (the path-TSP). However, this may not be the best solution for the firefighter: $L_\infty$~TSP minimizes the latest visit time for all fires, while the cost could be affected by all visit times; thus an aggregated measure could be a better objective.

For example, consider the following simple dynamics for the spread of wildfires over uniform territory, ignoring possible differences such as vegetation, weather and wind. After every second, a flammable point of territory is ignited if it is within a unit distance from the burning flames. Under these dynamics, the area of land scorched by the fire is a quadratic function of the elapsed time. Therefore, the damage due to the delay on the $i$\textsuperscript{th} visit can be better represented by ${\ell}_{\sigma_i}^2$ and minimizing $\sum_v \ell_v^2 = \|\ell\|_2^2$, or equivalently $\|\ell\|_2$, is a better objective for this scenario compared to other norms, particularly $\|\ell\|_1$ or $\|\ell\|_\infty$. Motivated by this example, we term $L_2$-TSP as the Traveling Firefighter Problem (TFP).

In an applied setting, this approach may require some refinement but the basic idea still applies. Land and weather asymmetries can be modeled by a multiplicity of vertices: If a fire spreads twice as fast in area, we can represent it by two vertices overlapping in the metric. One could also generalize the objective to a weighted sum of the squared delays and/or discretize large fires into smaller ones. Moreover, the time required to extinguish a fire can be accounted for by adding a new edge, hanging from the original destination at a distance proportional to the time required to contain that fire, and moving the destination to the other endpoint of the new edge. 

\begin{figure}[t]
    \centering
    \subfloat[$L_2$-TSP route]{{
    \begin{tikzpicture}
  \SetGraphUnit{2}
  \Vertex{S}
  \WE(S){A}
  \EA(S){B}
  \EA(B){C}
  \tikzset{EdgeStyle/.append style = {bend left}}
  \Edge(S)(A)
  \Edge(A)(B)
  \Edge(B)(C)
\end{tikzpicture}
    }}%
    \qquad
    \subfloat[$L_1$~TSP route]{{
    \begin{tikzpicture}
  \SetGraphUnit{2}
  \Vertex{S}
  \WE(S){A}
  \EA(S){B}
  \EA(B){C}
  \Edge(S)(B)
  \Edge(B)(C)
  \Edge(C)(A)
\end{tikzpicture}
    }}%
    \caption{Different norms lead to different optimal routes.}%
    \label{fig:L2vsL1TSP}
\end{figure}
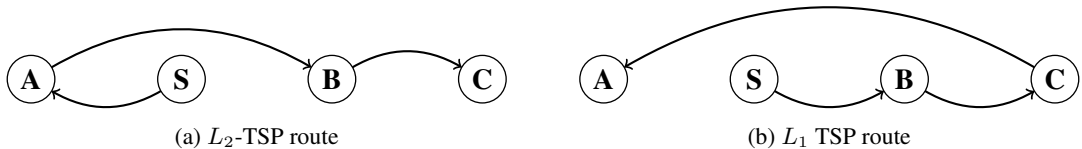

\begin{figure*}[t]
    \centering
    \subfloat[Path-TSP route]{{\includegraphics[trim={15cm 4.5cm 12cm 8cm},clip,width=6cm]{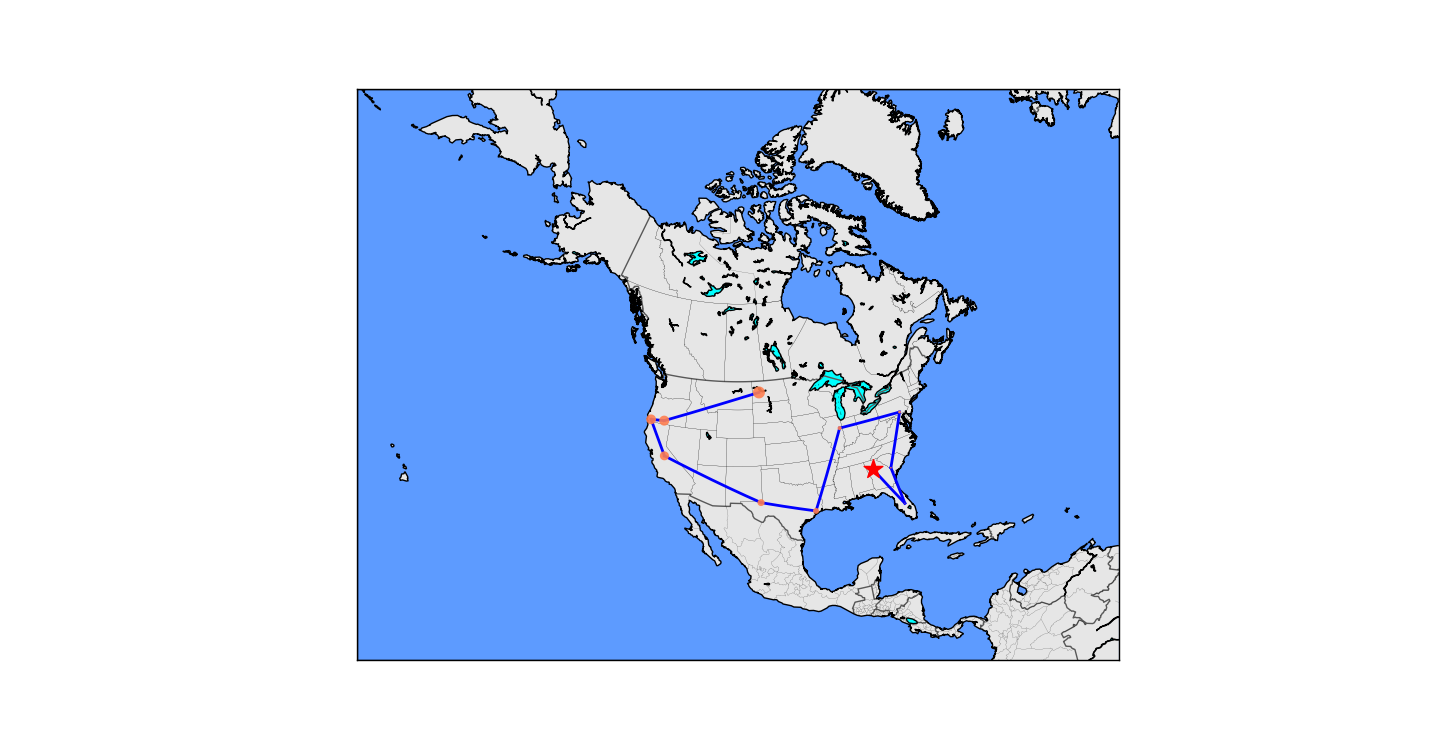}}}%
    \qquad
    \subfloat[TFP route]{{\includegraphics[trim={15cm 4.5cm 12cm 8cm},clip,width=6cm]{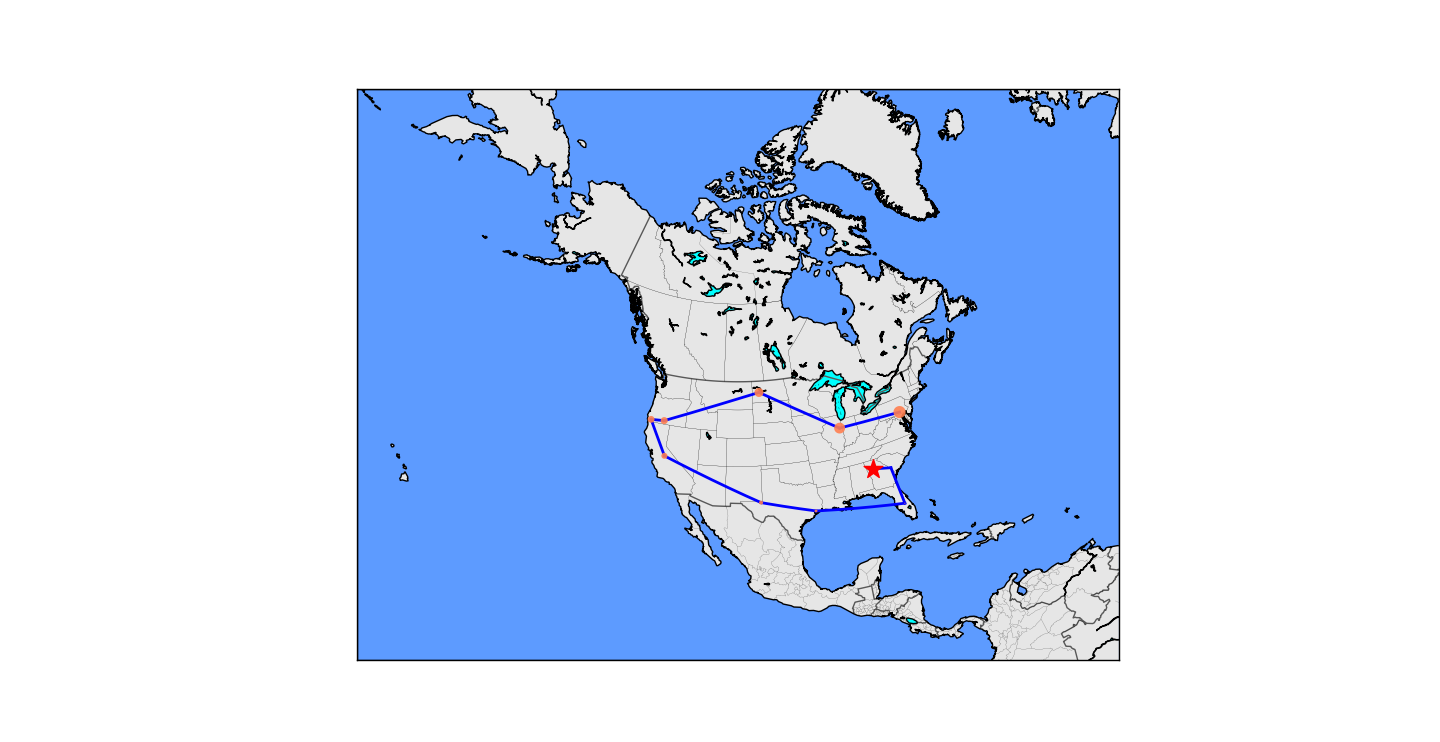}}}%
    \caption{Optimal path-TSP versus TFP routes for 10 locations in U.S.\ with simultaneous wildfires; the areas of the discs at visit locations are proportional to the expected (quadratic) damage at each location, due to the delay along the route. Here, taking the TFP route (instead of the TSP) reduces total damage by $5\%$.}%
    \label{US-routes}%
\end{figure*}

\noindent\textbf{The Argument versus The Objective.} The set of feasible routes for all $L_p$~TSP problems is the same, while the objectives are different. For any $p \neq q$, there exist instances where the two optimal routes are different. As a simple example, depicted by \Cref{fig:L2vsL1TSP}, consider four vertices \textbf{S}, \textbf{A}, \textbf{B}, and \textbf{C} over a line metric at locations $0,-1-\varepsilon,+1$, and $+2$ respectively. Starting at \textbf{S}, the route \textbf{SABC} is optimal for $L_2$-TSP with corresponding objective of $\|(0,1+\varepsilon, 3+2\varepsilon, 4+2\varepsilon)\|_2 \simeq \sqrt{26}$, in contrast to the optimal route for $L_1$~TSP that is \textbf{SBCA} with the objective $\|(0,1,2,5+\varepsilon)\|_1 = 8 + \varepsilon$.

Consider the effect of a ``wrong'' norm on the optimal route. \Cref{US-routes} demonstrates an example in which we consider $10$ locations with simultaneous wildfires in the United States.\footnote{Wildfires observed during the first week of Dec $2019$, according to satellite data \cite{NASA}.} Dispatching a firefighter from Atlanta, and traveling $100$ times faster than the spread of fire, TFP would look quite different than $L_\infty$~TSP, i.e., path-TSP. In particular, a TFP optimal route reduces the total damage by $5\%$ compared to that of the path-TSP.

As another example, consider destinations on the $x$ axis, with many fires located at $+1$ and a single fire at $-1+\varepsilon$. Starting at $x = 0$, the optimal $L_\infty$~TSP route moves left first and then right, resulting in a solution roughly three times as expensive as the optimal TFP route, in terms of the $L_2$ objective. In fact, the relative performance of an $L_\infty$~TSP route for the $L_2$ objective can be \emph{unbounded}. For instance, consider the Euclidean metric over the plane with $n-2$ fires located at complex coded locations $\{e^{2\pi \cdot \frac{k}{n} i} : k \in \{1, \cdots, n-2\}\}$, and $m$ distinct fires located at $e^{2\pi \frac{n-1-\varepsilon}{n} i}\,.$ Starting at $(1,0) = e^{0i}$ and moving at unit speed, for $n \rightarrow \infty$ and $m/n \rightarrow \infty$ the damage (squared delay) due to $L_\infty$~TSP route (by walking in the wrong direction around the circle) converges to $4\pi^2$, while for the optimal TFP solution it goes to zero.

\medskip
\noindent\textbf{All-Norm TSP.} 
We observed that TSP may not be a good solution for other $L_p$ objectives. One may consider whether a different $L_p$~TSP is hopefully good for all norms. In this line, a natural question is whether there exists a single route that is approximately optimal concerning the minimization of any norm of the visit times, and whether we can efficiently find one. This can be viewed also as an online problem where the adversary chooses the norm, e.g.\ $L_p$, and the objective is to provide a competitive solution with respect to the optimal route.

\medskip
\begin{definition}[All-Norm TSP]
Given $s, V \ni s, d: V \times V \rightarrow \mathbb{R}_{\ge 0}$ as before, the objective is to choose a route that minimizes the maximum possible ratio between a symmetric norm of the visit time vector of the output route $\sigma$ and the optimal route for that norm,
$
\min_{\sigma \in \mathcal{F}} \sup_{\| \cdot \|} \frac{ \lVert \ell_{\sigma} \rVert}{\min_{\sigma' \in \mathcal{F}} \lVert\ell_{\sigma'}\rVert}
$.
\end{definition}

Golovin et al.\ \cite{GGKT08} introduced this problem as the \emph{All-Norm TSP} and gave an algorithm that outputs a route that is a $16$-approximation with respect to any norm of the visit time vector. The concept of all-norm minimization has been of interest in many applications, e.g., for routing, load balancing \cite{KRT01}, and machine scheduling \cite{AERW04,BP03}.

\subsection{Summary of Results \& Techniques Overview}\label{sec:formal-results}
Optimizing the non-linear objective of $L_p$~TSP can be computationally challenging. The problem is strongly NP-hard even in the linear case, $p = 1$, on a tree metric \cite{S02}, i.e.\ when the metric is pairwise distances over a tree graph induced by $V$. In contrast, tree TSP is solvable in linear time. 

\noindent\textbf{Reduction to TSP.}
TSP is the most studied routing problem with various approximation algorithms on various metrics. A natural question is whether we can reduce $L_p$~TSP to (not many) instances of TSP? Note that a naive algorithm for $L_p$~TSP is to try all $n! = n^{\Omega(n)}$ candidate solutions. $L_p$~TSP can be reduced (preserving the approximation ration) into substantially fewer instances of TSP. We first need a result of Archer and Williamson \cite{AW03} who showed that a $(1+\varepsilon)$-approximate solution for TRP exists that is a concatenation of $O(\log n \cdot \varepsilon^{-1})$ TSP paths. This holds for $L_p$~TSP as well.

\begin{lemma}[\cite{AW03}]\label{lem:TSPsegments}
$L_p$~TSP can be $(1+\varepsilon)$-approximated by a concatenation of $O(\log n \cdot \varepsilon^{-1})$ TSP paths.
\end{lemma}

\begin{proof}
Without loss of generality we can assume integral and polynomially bounded input distances, $d(i,j) \in \set{0} \cup [O(n^2/\varepsilon)] = \set{0} \cup [\tilde{O}(n^2)]$,  as an appropriate quantization by rounding only adds a multiplicative error of $1 + O(\varepsilon/n^2)$ to any edge and keeps any norm of a valid tour, within a factor $(1 \pm \varepsilon)$. 

Now break the optimal route according to time spots $1, (1+\varepsilon), \cdots, (1+\varepsilon)^\gamma$ where $\gamma = O(\log n \cdot \varepsilon^{-1})$ because $O(n^3 \cdot \varepsilon^{-1})$ is a bound on the length of the optimal route. Replacing each sub-route between two consecutive time spots, with the shortest path TSP over the same points does not increase any visit time beyond a factor $1+\varepsilon$, and hence preserves any norm of the visit times by a factor of $1+\varepsilon$.
\end{proof}

\begin{observation}\label{obs:TSPreduction}
$L_p$~TSP can be reduced to $O(\log n \cdot \varepsilon^{-1})^n$ many instances of path-TSP (or equivalently TSP \cite{TVZ20}), preserving the approximation factor within $1+\varepsilon$.
\end{observation}

\begin{proof}
Due to \Cref{lem:TSPsegments} there exists a $1+\varepsilon$ approximate solution for $L_p$~TSP composed of $t = O(\log n \cdot \varepsilon^{-1})$ TSP paths, namely $t$ segments. We can (partially) recover this approximately optimal solution by enumerating over $t^n$ all such candidates. In each case compute the underlying metric induced by vertices of each segment, i.e., with the distance between two vertices being the length of the shortest path in between due the original metric. Calling an $\alpha$-approximation oracle for each segment and concatenating the resulting segments, report the best such feasible solution (with respect to the $L_p$~TSP objective) as the answer.
\end{proof}

The above approach can lead to a quasi-polynomial time approximation algorithm for trees, and remains (super) exponential for general metrics. In fact, reducing the problem to many shortest paths (TSP) cannot lead to an efficient polynomial-time (approximate) reduction, because a concatenation of $o(\log n)$ path-TSP routes cannot approximate $L_1$~TSP within a constant factor \cite{S14}. To further reduce the number of paths, we use the notion of segmented-TSP, introduced by Sitters \cite{S14}, to enable some dependence between consecutive sub-routes. This problem requires a (sequence of monotonically non-decreasing) number of destinations to be visited by a number of deadlines, formulated as follows.

\begin{definition}[segmented-TSP]
Given $V \ni s, d: V \times V \rightarrow \mathbb{R}_{\ge 0}$ as before, in addition to integer numbers $n_1 \leq n_2 \leq \cdots \leq n_k \leq |V| = n$ and fractional numbers $t_1 \leq t_2 \leq \cdots \leq t_k$ as inputs, the segmented-TSP problem is a decision problem to verify whether a route exists that visits at least $n_i$ distinct vertices by time  $t_i$ for all $i \in [k]$, starting at $s$. 
\end{definition}

Approximation of segmented-TSP, i.e.\ a decision problem, can be defined as follows.

\begin{definition}
An $\alpha$-approximate solution to a segmented-TSP instance, must visit the first $n_i$ vertices by the modified deadline $\alpha \cdot t_i$, $\forall i \in [k]$, if an answer to the original segmented-TSP exists.
\end{definition}

We generalize a main result of Sitters \cite{S14} that showed TRP can be reduced to (a polynomially many number of approximate) segmented-TSP problems with a constant number of deadlines.

\begin{theorem}
\label{thm:reduction}
Let $\varepsilon > 0$ be a constant and $\mathcal{A}$ be an $\alpha$-approximation algorithm for segmented-TSP for some $k = O(1+\varepsilon^{-2})$ of our choice. There is a $(1+\varepsilon) \cdot \alpha$-approximation algorithm for $L_p$~TSP that calls $\mathcal{A}$ (on the same network $(V,s,d)$) for a strongly polynomial number of times.
\end{theorem}

\Cref{thm:reduction}, proved in \Cref{sec:complx}, along with a PTAS for segmented-TSP \cite{S14} on tree metrics and Euclidean metrics imply the following results.

\begin{restatable}{corollary}{PTAScorollary}\label{thm:PTAS}
There exist polynomial-time approximation schemes (PTAS) for $L_p$~TSP on weighted trees and Euclidean plane metrics.
\end{restatable}

Note that due to our reduction of $L_p$~TSP to segmented-TSP, any approximation results for segmented-TSP, on specific metrics, would follow for $L_p$~TSP, at the cost of an additional factor of $(1+\varepsilon)$ to the approximation bound. Our results can be generalized to the case of multiple travelers, starting from arbitrary locations, as discussed in \Cref{sec:multi}.

For general metrics, a PTAS is unlikely, as the problem becomes Max-SNP-hard. On the other hand, the constant factor approximability of $L_p$~TSP for general metrics is immediate due to the $16$-approximation of the All-Norm TSP by Golovin et.\ al.\ \cite{GGKT08}. We improve the approximation bound for All-Norm TSP by a factor of $2$.

\begin{theorem}\label{thm:8apx}
There is a polynomial-time algorithm to find a route that is $8+\varepsilon$-approximate with respect to the minimization of any symmetric norm of the visit times (including $L_p$~TSP, TFP, and TRP).
\end{theorem}

\Cref{thm:8apx} is proved in \Cref{sec:all-norm}.
Our algorithm builds on the partial covering idea that was pioneered by Blum et al.\ \cite{BCCPRS94} for TRP, and was developed through subsequent studies \cite{GK98, CGRT03, GGKT08,bienkowski2021traveling}. It is presented as \Cref{meta-algorithm}.

\makeatletter
\def\BState{\State\hskip-\ALG@thistlm}
\makeatother

\begin{algorithm}
\caption{Routing via Partial Covering}\label{meta-algorithm}
\begin{algorithmic}[1]
\Procedure{Geometric-Covering}{$V, s, d$}
\State Algorithm Parameters: $b \in (0,\infty),c \in (1,\infty)$ 
\For{$i \gets 0, 1, 2, \dots$} \Comment{Conducting sub-tours, as long as un-visited destinations remain}
        \State $C_i \gets$ a maximal route of length $\leq b \cdot c^i$. 
        \State Travel through $C_i$ (and return to the origin)
\EndFor
\State \Return an ordering $\sigma$ of $V$ according to their (first) visit time through the above loop.
\EndProcedure
\end{algorithmic}
\end{algorithm}

The (hyper) parameters $b$ and $c$ strongly affect the performance of the algorithm. For All-Norm TSP, we choose $b = \min_{i,j \in V} d(i, j) $ and $c = 2$, as in \cite{GGKT08}. The key difference of our algorithm is in line $4$. Instead of an approximation algorithm for $k$-TSP, i.e.\ a route of minimum length that visits $k$-vertices, we utilize a milder relaxation, which is a tree (instead of a route/stroll) rooted at $s$, including $k$ vertices, and of total length not larger than an optimal $k$-TSP. Such a \emph{good $k$-tree} can be $1+\varepsilon$ approximated in polynomial time using the primal-dual method that solves a Lagrangian relaxation of $k$-TSP \cite{CGRT03,ALW08,PS14}.

On the other hand, we provide a $1.78$ approximation impossibility result for All-Norm TSP, notably beyond any known inapproximability bounds for $L_p$~TSP problems. The following is proved in \Cref{sub:inappx}.

\begin{theorem}\label{thm:allnormlowerbound}
There is no approximation algorithm for All-Norm TSP with multiplicative factor better than $1.78$, independent of $P = NP$ or other complexity hypotheses.
\end{theorem}

The above result reaffirms the need for approximation algorithms specifically designed for each norm. Along this line, we present a randomized approximation algorithm for the Traveling Firefighter Problem.
\begin{theorem}\label{thm:TFP}
There is a randomized, polynomial-time $4.27$-approximation algorithm for TFP on general metrics.
\end{theorem}

\Cref{thm:TFP} is proved in \Cref{sec:TFP}, for which we build upon the ideas by Chaudhuri et.\ al.\ \cite{CGRT03}. We analyze \Cref{meta-algorithm} where
$b \in [1,c]$ is chosen at random, such that $\log(b)$ uniformly distributed, to simplify the analysis while one can efficiently de-randomize the algorithm by quantization of $b$. The key is to analyze the hyper parameter $c$ to minimize the approximation bound. Same approach for general $L_p$~TSP led to the following.

\begin{theorem}\label{thm:LPTSP}
$L_p$~TSP can be $8/(p \ln 4)^{1/p}$ approximated in polynomial time. 
\end{theorem}

A key step of our combinatorial algorithms for the above results is line $4$ for which we use \emph{good $l$-trees}. An $l$-path is defined as a (simple) path that visits at least $l$ vertices. A good $l$-tree is a tree with at least $l$ vertices, with a length of at most that of the minimum-length $l$-path. Chaudhury \emph{et al.} \cite{CGRT03} showed good $l$-trees can be found in polynomial time. However, generalizing the theory of good-$l$-tree to multi-vehicle case (in order to extend \Cref{meta-algorithm} for multi-vehicle $L_p$~TSP) is not straightforward. The good $l$-tree problem is also related to the path orienteering problem and the budgeted prize-collecting TSP problem \cite{archer2011improved, chekuri2012improved, paul2020budgeted}.

We re-prove \Cref{thm:TFP} and \Cref{thm:LPTSP} using a linear programming rounding approach. We generalize a natural linear programming relaxation of TRP \cite{CS11, PS14} to a relaxation for $L_p$~TSP and present randomized rounding algorithms with matching approximation guarantees to that of the combinatorial algorithm. An advantage of this method, as we shall see, is extendability to
multi-vehicle problems thanks to independence of corresponding constraints for different vehicles in the dual linear program. We describe the linear programming relaxation (and its solvability) in \Cref{sec: lp_solution} and present the rounding algorithm and analyze it in \Cref{sec: rounding_algorithm}. Our theory ensures the following approximation bounds for multi-vehicle $L_p$~TSP problems.

\begin{theorem}\label{thm:lp_intro_master_theorem_multi_vehicle}
Multi-vehicle TFP and multi-vehicle $L_p$~TSP can be approximated by factors $10.92$ and $18p$, respectively, in polynomial time.
\end{theorem}

The above is restated and proved as \Cref{thm: lp_master_theorem_multi_vehicle}. Before we end this Section, let us summarize our key results in \Cref{tab:routing}.

\begin{table}[t]
\footnotesize
\centering
 \caption{Polytime Approximation of Multi Purpose Routing Problems.} \label{tab:routing}
\begin{tabular}{||c || c | c | c||} 
\hline
 Problem & Polytime Approximation & Reference & Previous Best \\ [0.5ex] 
 \hline\hline
 $L_2$-TSP & $4.27$ &  {\Cref{thm:TFP}} (comb.) \& {\Cref{thm: lp_master_theorem_single_vehicle}} (LP) & 5.65 \cite{FTT21}  \\
 $L_p$-TSP & $8/(p \ln 4)^{1/p}$ & {\Cref{thm:LPTSP}} (comb.) \& {\Cref{thm: lp_master_theorem_single_vehicle}} (LP) & - \\
 \hline \hline
 All-Norm TSP & $8+\varepsilon$ (\& $1.78$ impossible) & {\Cref{thm:8apx}} \& \Cref{thm:allnormlowerbound} & 16 \cite{GGKT08} \\
 \hline \hline
 Tree \& Euclidean $L_p$ TSP & $1+\varepsilon$ & {\Cref{thm:PTAS}} & $L_1$-TSP \cite{S14} \\
 \hline \hline
 Multi Vehicle/Depot $L_2$ TSP & $10.92$ & \multirow{2}{*}{{\Cref{thm:lp_intro_master_theorem_multi_vehicle}}} & - \\
 Multi Vehicle/Depot $L_p$ TSP & $18p$ & & - \\ 
 \hline
\end{tabular}
\end{table}

\section{Reducing $L_p$~TSP to Segmented TSP}\label{sec:complx}

In this section, we provide a reduction from $L_p$~TSP to polynomially many instances of segmented-TSP, i.e., proving \Cref{thm:reduction}.
In particular, a corollary of this result (along with the following Lemma) is a $(1+\varepsilon)$ approximation algorithm for any $L_p$~TSP on weighted-tree metrics -- where the problem becomes strongly NP-hard even for $p = 1$ -- as well as the Euclidean plane. 

\begin{lemma}[\cite{S14}]
\label{segTSP}
Segmented TSP, for any constant number of segments $M$, can be solved in polynomial time for weighted trees, and $1+\varepsilon$ approximated for unweighted Euclidean metric.
\end{lemma}

\PTAScorollary*

In the rest of this section, we prove \Cref{thm:reduction}, by providing a dynamic programming algorithm that approximates $L_p$~TSP using polynomially many calls to (approximate) segmented-TSP. More precisely, we show that given an $\alpha$-approximate solver for segmented-TSP, the dynamic program guarantees an approximation factor of at most $\alpha\cdot(1+\varepsilon)$ for arbitrary constant $\varepsilon > 0\,.$

The algorithm is presented in a few steps, each imposing no more than $1+O(\varepsilon)$ multiplicative error. To achieve exact $1+\varepsilon$ precision, one may run the algorithm for a constant fraction of the target $\varepsilon\,.$ 

For some $k$ as large as $ O(1+\varepsilon^{-2})$ we can ensure $c \defeq (1+\varepsilon)^k \ge 3\,.$ Let $\Opt^{\lambda_i}$ denote the maximal prefix of $\Opt$ route for $L_p$~TSP of length at most
$
\lambda_i \defeq (1+\varepsilon)^{-j} \cdot c^i\,, \quad \forall i \ge 0
$,
where $j$ is a fixed random number, uniformly distributed over $\{0, \cdots, k-1\}$.

Let $\Opt'$ be a tour made of sub-tours consisting of traversing $\Opt^{\lambda_i}$ and returning to the origin and waiting until time $3\lambda_i$ before starting the next sub-tour. 
To confirm the above is feasible, we need to show sub-tour $i+1\,,$ being allowed to begin at $3\lambda_i\,,$ does not leave before the return of previous sub-tour, i.e.,
$
3\lambda_{i-1} + 2 \|T^{\Opt^{\lambda_i}}\|_\infty \leq \lambda_i + 2\lambda_i = 3\lambda_i
$
which is immediate having $\lambda_i = c \lambda_{i-1} \ge 3\lambda_{i-1}$.

\begin{lemma}\label{lem:loss}
The modified tour $\Opt'$ is approximately optimal in expectation, i.e.,  for some $k \in O(1+\varepsilon^{-2})$ we will have $\expec{j}{\|T^{\Opt'}\|_p^p} \leq (1+\varepsilon) \|T^\Opt\|_p^p$.
\end{lemma}

\begin{proof}
All vertices are visited (for the first time) in the same order, by $\Opt$ and $\Opt'\,.$
Let the $d$\textsuperscript{th} service time by the optimal solution be $T_d^\Opt \in ((1+\varepsilon)^\delta, (1+\varepsilon)^{\delta+1}]$ for some integer $\delta \ge 0\,.$
If this vertex is visited in the $i$\textsuperscript{th} sub-tour of $\Opt'\,,$ we can write
$
T_d^{\Opt'} =  T_d^{\Opt} + 3\lambda_{i-1}\,.
$
We can bound this additional delay by 
$T_d^{\Opt'} - T_d^{\Opt} \leq 3(1+\varepsilon)^{\delta-j'}$ where $j'$ has the same distribution as $j\,.$
We can prove the desired by bounding the per-vertex ratio by
$
{\expec{j'}{(T_d^{\Opt'})^p}}/{(T_d^\Opt)^p}
 \leq (1+\varepsilon)
$
because
$
\frac{\sum_{i} a_i}{\sum_{i} b_i} \leq \max_{i} \frac{a_i}{b_i}
$
where $a_1, \cdots$ and $b_1, \cdots$ are positive real numbers.

Considering $p \ge 1$ and $T_d^{\Opt'} = T_d^{\Opt} + 3\lambda_{i-1}$, the left hand side of the target ratio is maximized for $T_d^{\Opt} = (1+\varepsilon)^\delta$, so it suffices to prove
$
\frac{\expec{j'}{((1+\varepsilon)^\delta + 3\lambda_{i-1})^p}}{((1+\varepsilon)^\delta)^p}
 \leq (1+\varepsilon)\,.$
We will have
\begin{align*}
\frac{\expec{j'}{((1+\varepsilon)^\delta + 3\lambda_{i-1})^p}}{((1+\varepsilon)^\delta)^p} 
&\leq
\frac{\expec{j'}{((1+\varepsilon)^\delta + 3(1+\varepsilon)^{\delta-j'})^p}}{((1+\varepsilon)^\delta)^p} 
=
{\expec{j'}{(1 + 3(1+\varepsilon)^{-j'})^p}} \\
&=
1 + {\expec{j'}{(1 + 3(1+\varepsilon)^{-j'})^p- 1^p}} 
\leq
1 + \frac{1}{k} \sum_{j' = 0}^{k-1} (3p)^p (1+\varepsilon)^{-j'} \\
&\leq
1 + \frac{(3p)^p}{k} \cdot \frac{1}{1-(1+\varepsilon)^{-1}} 
=
1 + \frac{(3p)^p}{k} \cdot \frac{1+\varepsilon}{\varepsilon}\,.
\end{align*}

To get the desired (from the last inequality) it suffices to assume
$
k \ge \frac{(3p)^p(1+\varepsilon)}{\varepsilon^2}\,.
$
\end{proof}

We can now complete the proof of \Cref{thm:reduction}, generalizing the main result of Sitters \cite{S14}.

\Cref{lem:loss} implies for some $j \in [k]$, where $k = O(1+\varepsilon^{-2})$, there exists a near optimal routing, $\Opt'\,,$ that for each $i \in [\tilde{O}(n^2)]$, visits \emph{new vertices} only during $[3\lambda_{i-1},\lambda_i]\,,$ and returns to the origin and remains there until $3\lambda_i = \frac{3}{(1+\varepsilon)^j} \cdot (1+\varepsilon)^{ki}\,.$ We can search for such a path by reconstructing $\Opt^{\lambda_i}$ for all $i$ and upper bounding the consequent $\|T^{\Opt'}\|_p^p$ using dynamic programming.

Define $D[i][d]$ as (an upper bound on) the contribution of visit times of vertices that are visited by $\Opt^{\lambda_i}$, to $\|T^{\Opt'}\|_p^p$, further assuming the number of these vertices is $d$. We can compute $D[i][d]$ considering $O(n^k)$ cases of $(m_1, m_2, \cdots, m_k)$ where $m_r$ denotes the number of vertices that are visited by $\Opt^{\lambda_i}$ during $(3\lambda_{i-1} + \lambda_i \cdot (1+\varepsilon)^{r-k-1}, 
3\lambda_{i-1} + \lambda_i \cdot (1+\varepsilon)^{r-k}]$. Note that it is necessary to have $\sum_r m_r \leq d$. Let $d' = d - \sum_r m_r$ be the number of vertices visited by $\Opt^{\lambda_{i-1}}$. We can write
\[
D[i][d] = \min_{m_1,\cdots,m_k} D[i-1][d'] + \text{Seg-TSP}_i(d',m_{[r]}) \cdot \sum_r m_r \cdot (3\lambda_{i-1} + \lambda_i \cdot (1+\varepsilon)^{r-k})^p\,,
\]
where $\text{Seg-TSP}_i(d',m_{[r]})$ has value $1$ if segmented-TSP is feasible for visiting at least $d',d'+m_1,\cdots,d'+m_1+\cdots+m_r$ vertices by deadlines $\lambda_{i-1},3\lambda_{i-1}+\lambda_i \cdot (1+\varepsilon)^{-k}, \cdots, 3\lambda_{i-1} + \lambda_i$, respectively, is feasible. Otherwise let $\text{Seg-TSP}_i(d',m_{[r]}) = \infty$. Note that we have an $\alpha$ approximate solver for Segmented-TSP, though for convenience we can alternatively assume the traveller goes at the speed of $\alpha$ instead of $1$, to get a $1+\varepsilon$ approximate solution to $\|T^{\Opt}\|_p$ by $(D[\tilde{O}(n^2)][n])^{1/p}$. In the end, moving at unit speed (instead of $\alpha$) at every stage can increase (any) norm of the delay vector $\|T\|$ by a factor $\alpha$ so we have an $\alpha \cdot (1+\varepsilon)$ approximation, that was promised by \Cref{thm:reduction}.

Finally it is worth to mention that the route (instead of the value) can be reconstructed using update (parent) information of $D[\cdot][\cdot]$ and a constructive approximate solver for $\text{Seg-TSP}[\cdot]$ and we can short-cut potential re-visits of vertices to have a valid Hamiltonian route.

\section{ All-norm TSP}\label{sec:all-norm}

Since the introduction of a first constant approximation for TRP by Blum et al.\ \cite{BCCPRS94}, partial covering through applying a geometric series of limits on the length of the sub-tours has been a core in the design of routing algorithms. In this section, we improve the $16$-approximate/competitive solution for all-norm TSP by a factor of $2$. We further provide a first lower bound for this problem.

\subsection{$8$ Approximation for General Metrics.}

The idea is to iteratively cover more and more vertices by sub-tours of exponentially increasing length while trying to maximize the total number of vertices that are visited (not necessarily for the first time) in each iteration. Our algorithm uses the following milder relaxation of $k$-TSP, called a \emph{good $k$-tree}, in place of line $6$ in \Cref{meta-algorithm}.

\begin{definition}
A good $k$-tree is a tree of size $k$, including $s$, and with a total edge-weight of no more than that of the optimal $k$-TSP (starting from $s$).
\end{definition}

\begin{lemma}[\cite{CGRT03}]\label{lem:goodktree}
A good $k$-tree can be $1+\varepsilon$ approximated in polynomial time.
\end{lemma}

Chaudhuri et.\ al.\ \cite{CGRT03} proved the above using a primal-dual approach \cite{garg19963,AK00} that allows finding a feasible solution to the primal (integer) linear program of the $k$-tree problem paired with a feasible dual solution to $k$-TSP, that by weak duality has no less of a cost. 

We are now ready to prove \Cref{thm:8apx}.

\begin{proof}[Proof of \Cref{thm:8apx}]
WLOG assume the nearest neighbor to $s$ is at distance $1$, and for simplicity assume we have exact good $k$-trees which adds a multiplicative $1+O(\varepsilon)$ to the approximation bound. For $k = 1, 2, \dots, n$ find a good-$k$-tree. Among these, name the largest tree (with respect to number of vertices) of total length at most $2^i$ as $G_i$ for $i = 0, 1, 2, \dots\,.$

Let $C_i$ be a (randomized) depth-first traversal of $G_i$ and let $C$ be the concatenation of $C_i$'s for $i = 0, \dots\,.$ The final tour $\Alg$ will visit vertices in the order that they appear in $C$, which does not increase the first-visit time for any vertex, due to triangle inequality of the metric, while short-cutting vertices that are being re-visited. 

Let $$T_k^\Opt \in [2^i, 2^{i+1})\,.$$ This shows the shortest (length) $k$-path in $G$ is no longer than $2^{i+1}\,.$ So the good-$k$-tree is no longer than $2^{i+1}\,,$ hence $C_{i+1}$ has at least $k$ distinct vertices, allowing us to upper bound the $k$\textsuperscript{th} visit time by
$$T_{k}^\Alg \leq \sum_{j = 0}^{i+1} |C_j| \leq \sum_{j = 0}^{i+1} 2 \times 2^{j} < 2^{i+3}\,.$$

Together with $T_k^\Opt \ge 2^i$ and the above inequality we have
$$
T_k^\Alg \leq 8 \times T_k^\Opt\,.
$$

We showed
$T_k^\Alg \leq 8 \cdot T_k^\Opt \quad \forall i \in [n]$, i.e., $T^\Opt$ is $8$-submajorized by $T^\Alg$ in terminology of \cite{GGKT08,HLP88}. The result is that $$\|T^\Alg\| \leq 8 \cdot \|T^\Opt\|$$ w.r.t.\ any norm $\|\cdot\|$.
\end{proof}

One can verify the above algorithm performs asymptotically $3$ times worse than the optimal TRP for the example with service points at $\{2^i : i \in \mathbb{N}\}$ and starting at $x = 0\,.$

\subsection{A Factor $1.78$ Inapproximability.}\label{sub:inappx}
We conclude this section by providing a lower bound for all-norm TSP. We show even for line metrics, an $\alpha$-approximate all-norm TSP cannot be guaranteed in general, for $\alpha < 1.78$.

\begin{figure*}
    \centering
    {\includegraphics[width=15cm]{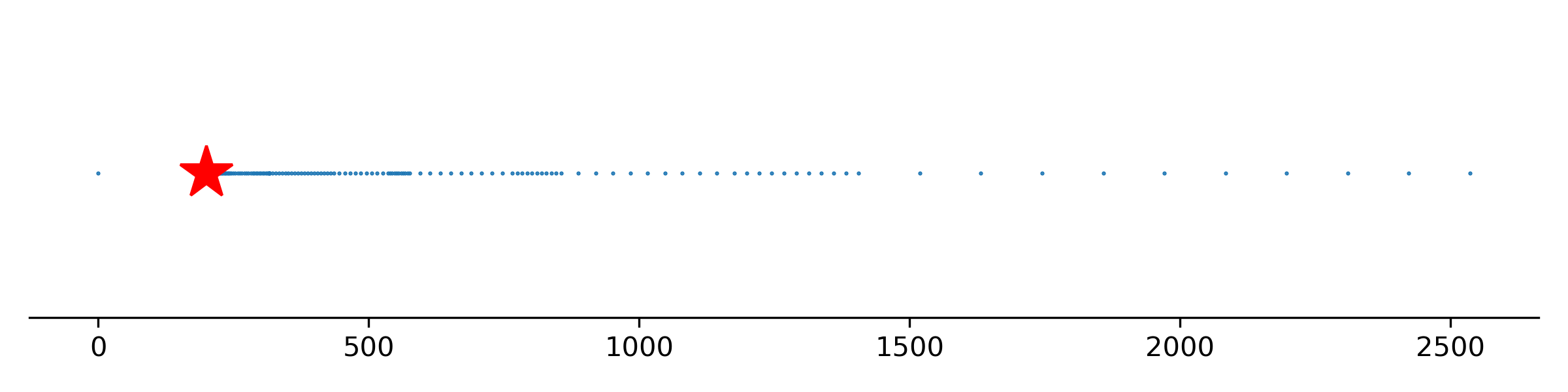}}
    \caption{The example with no better than $1.78$ all-norm TSP. The locations to visit are marked in blue, and the starting position is $x = 0$ (see Appendix \ref{sec:inapprox-example} for details).}%
    \label{fig-example}%
\end{figure*}

\begin{figure}[t]
    \centering
    \includegraphics[width=0.8\textwidth]{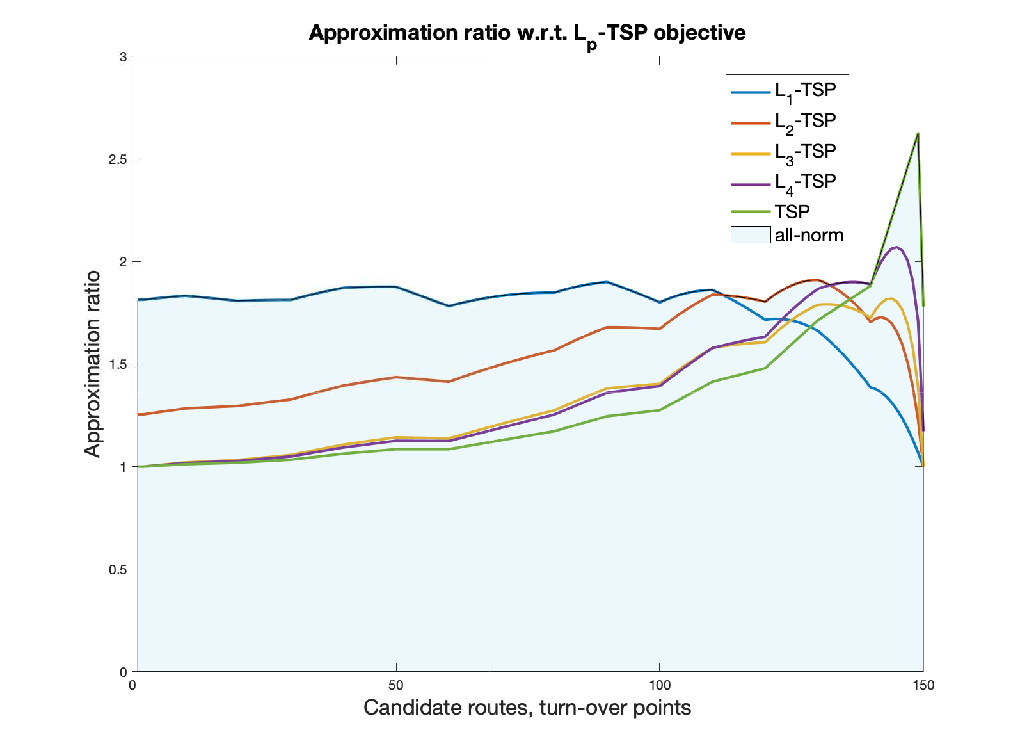}
    \caption{Proof by picture: Nonexistence of a $1.78$-approximate all-norm-TSP.}
    \label{fig:last}
\end{figure}

\begin{proof}[Proof of \Cref{thm:allnormlowerbound}]
We prove this for the special case of a line metric, i.e., when distances are absolute differences between points (vertices) on the real line. Consider the following simple example first: starting the walk from the origin at $x = 0$, there is a single destination at $x = -1$, in addition to $n$ destinations at $x \in \set{b^i-1 : i \in [n]}$, for $b = 1 + \varepsilon$. The approximation ratio of the route that first visits points to the right of the origin with respect to $L_{\infty}$-TSP objective is
$
\frac{2b^{n}-1}{b^{n}+1}
$,
that converges to $2$ when $n \rightarrow \infty$.
Alternatively, the approximation ratio w.r.t. $L_1$~TSP for the route that first goes left, i.e., optimal $L_{\infty}$ route, is
$
\frac{1+2n+b^{n+1}/(b-1)-n-1}{b^{n+1}/(b-1)-n-1+2b^n-1}\,.
$
The minimum of these two ratios will be at least $1.67$, achieved for $n = 2100$, $\varepsilon = 10^{-3}$.

We can construct a numerical example with similar structure, depicted in Figure \ref{fig-example}. We considered the approximation ratios with respect to various norms for all candidate optimal routes. The $\min \max$ over these ratios was $1.78$, verifying nonexistence of an approximate All-Norm TSP with better performance. We include this example in \Cref{sec:inapprox-example}, for which Figure \ref{fig:last} depicts performance of candidate (for all-norm optimality) routes with respect to different norms.

\end{proof}

The above example, is yet another motivation to study and optimize routing algorithms specific to the appropriate objective/norm, as one solution \emph{cannot} be good for all.

\section{Optimal Routing of a Firefighter: a Combinatorial Algorithm}\label{sec:TFP}

In this section we build upon the geometric partial-covering algorithm with good $k$-trees to improve approximation bound for a specific norm, i.e., the Traveling Firefighter Problem.

We achieve the improved approximation bound by randomization (of parameter $b$) and optimization of the analysis w.r.t.\ approximation guarantee for a specific norm. This approach can provide approximation guarantees (better than $8$) for other $L_p$~TSP problems. We present main ideas in the rest of the section by proving \Cref{thm:TFP}.

\begin{proof}[Proof of \Cref{thm:TFP}]
We present a randomized approximation algorithm for general metrics, that can be efficiently de-randomized.

Among the set of good $k$-trees, pre-computed for all $k$ using \Cref{lem:goodktree}, let $G_i$ be the largest one of total length at most $b\cdot c^i\,,$ and let $C_i$ be a depth first traversal of that. Parameter $c > 1$ is a constant, to be optimized for performance guarantees, and $[1,c] \ni b = c^U$ where $U$ is a random variable distributed uniformly over the interval $[0,1]\,.$ Finally, we reverse each $C_i$ with probability half, and concatenate $C_i$'s (and shortcut repeated visits) to achieve the output ordering $\Alg\,.$
Let the latency of the $k$\textsuperscript{th} vertex visited by the optimal route be $T_k^\Opt = a c^i\,,$ for some $a \in [1, c]$ and integer $i \ge 0\,.$ 

It is easy to see that the $(i + \mathbbm{1}[a \ge b])$\textsuperscript{th} sub-tour contains at least $k$ vertices, hence, we can bound
$$
T_k^\Alg \leq X_k + \sum_{j = 0}^{i - \mathbbm{1}[a < b]} 2 b c^j\,,
$$
where $X_k$ is zero if the tour visits its $k$\textsuperscript{th} vertex before the $(i - \mathbbm{1}[{a < b}])$\textsuperscript{th} sub-tour, and $X_k \in [0, 2bc^{i + \mathbbm{1}[a \ge b]}]\,,$ depending on its location in the sub-tour.
We can bound the expected damage, by
\begin{align*}
\expec{}{(T_k^\Alg)^2} &\leq \frac{1}{2} \left( \sum_{j = 0}^{i - \mathbbm{1}[a < b]} 2 b c^j \right)^2 \\
&+ \frac{1}{2} \left(2bc^{i+1 - \mathbbm{1}[a < b]} + \sum_{j = 0}^{i - \mathbbm{1}[a < b]} 2 b c^j\right)^2 \\
&= \frac{1}{2}\left( 2b \frac{c^{i + \mathbbm{1}[a \ge b]}-1}{c-1} \right)^2 + \frac{1}{2}\left( 2b \frac{c^{i+1+ \mathbbm{1}[a \ge b]}-1}{c-1} \right)^2\\
&\le \frac{1}{2}\left( 2b \frac{c^{i + \mathbbm{1}[a \ge b]}}{c-1} \right)^2 + \frac{1}{2}\left( 2b \frac{c^{i+1+ \mathbbm{1}[a \ge b]}}{c-1} \right)^2\\
&= \frac{1}{2} \frac{(2b)^2{(c^2 + 1)}}{(c-1)^2} c^{2(i+\mathbbm{1}[a \ge b])} \\
&= c^{2i} \cdot \frac{2{ (c^2 + 1)}}{(c-1)^2} \Bigl(b^2c^{2 \cdot \mathbbm{1}[a \ge b]}\Bigr)\,.
\end{align*}
Bounding the expected damage at the $i$\textsuperscript{th} service, considering random variable $b$ we have
\begin{align*}
\expec{}{(T_k^\Alg)^2}  &\leq \expec{}{c^{2i} \cdot \frac{2(c^2 + 1)}{(c-1)^2} \left(b^2c^{\mathbbm{2}[a \ge b]}\right)} \\
&= c^{2i} \cdot \frac{2(c^2 + 1)}{(c-1)^2} \left( c^2\int_0^{\log_c{a}} c^{2U} dU  + \int_{\log_c{a}}^1 c^{2U} dU \right) \\
&= c^{2i} \cdot \frac{2(c^2 + 1)}{(c-1)^2} \left( c^2 \Big[ \frac{c^{2U}}{\ln c^2}\Big] \bigg|_{0}^{\log_c{a}}  + \Big[ \frac{c^{2U}}{\ln c^2}\Big] \bigg|_{\log_c{a}}^{1} \right) \\
&= c^{2i} \frac{2(c^2 + 1)}{(c-1)^2} \left( \frac{c^2(a^2-1)}{{2 }\ln c} + \frac{c^2 - a^2}{{ 2}\ln c}\right) \\
&= (ac^i)^2 \left(\frac{2(c^2 + 1) (c^2-1)}{{ 2}(c-1)^2 \ln c}\right) \\
&= (T_k^O)^2 \left(\frac{(c^2 + 1)(c+1)}{(c-1)\ln c}\right)\,.
\end{align*}

We can now choose $c$ in order to minimize the multiplicative bound
\[
    \frac{(c+1)(c^2 + 1)}{(c-1) \ln c} \leq 18.154 + \delta'
\]
for an arbitrary $\delta' > 0$, that can be achieved for  $c = e - \epsilon$ for a small enough $\epsilon > 0$. With this we have
\[
\expec{}{\|T^\Alg\|_2^2} \leq (18.154 + \epsilon') \|T^\Opt\|_2^2\,,
\]
hence
$
    \expec{}{\|T^\Alg\|_2} \leq (4.261 + \epsilon) \|T^\Opt\|_2\,.
$
for an arbitrary $\epsilon > 0$.
\end{proof}

The above algorithm can be de-randomized by exploring all values for a dense enough quantization of $b\,,$ or be simply used by repeating the randomized algorithm and reporting the best route.

\subsection{Generalizing the Combinatorial Algorithm for $L_p$~TSP}\label{sec:combinatorial_algorithm}

The combinatorial algorithm from \Cref{sec:TFP}, stated as \Cref{alg: combinatorial-tfp}, can be analyzed for arbitrary $L_p$~TSP; which for completeness we provide in this Section along with a proof of \Cref{thm:LPTSP}.

\begin{algorithm}[H]
\caption{Routing by Partial Covering: A Combinatorial Algorithm}\label{alg: combinatorial-tfp}
    \begin{algorithmic}[1]
    \State Let $c > 1$ be a constant that we fix later, and $b = c^U$, where $U \sim U[0, 1)$, that is, $U$ is sampled uniformly from $[0, 1)$. 
    \For {$j = 0, 1, \ldots $,}:
        \State  - Using \Cref{lem:goodktree}, get a tree $T_j$ of length at most $b c^j$ covering most vertices
        \State - Double $T_j$ to get a cycle $C_j$, short-cutting any vertices already visited
        \State - Traverse $C_j$ from $s$ in either direction with probability $1/2$ to get a tour $Z_j$
        \State - Exit if every vertex has been covered by some tour
    \EndFor
    
    \State Concatenate the subtours $Z_{0}, Z_1, \ldots$ to obtain the final tour $Z$
    \State \Return an ordering $\sigma$ of $V$ according to their (first) visit time in $Z$
    \end{algorithmic}
\end{algorithm}

\begin{theorem}\label{thm: combinatorial_approximation} \Cref{alg: combinatorial-tfp} is a $\frac{8}{(p \ln 4)^{1/p}}$-approximation to $L_p$~TSP for all $p \ge 1$.
\end{theorem}

\begin{proof}
    Let the latency of the $l$\textsuperscript{th} vertex visited by the optimal route be
    \[
        T_l^{OPT} = a c^j
    \]
    for some $a \in [1, c)$ and integer $j \ge 0$. Let $T_l$ denote the latency of the $l$\textsuperscript{th} vertex visited by the algorithm. Following an argument similar to that in \Cref{thm: lp_master_theorem_single_vehicle}, it is enough to show that for all $l \in \big[ |V| \big]$,
    \[
        \mathbb{E} \big[ T_l^p\big] \le C \big(T_l^{OPT}\big)^p
    \]
    for an appropriate constant $C$ to show that the algorithm is a $C^{1/p}$ approximation.

    Note that $T_l^{OPT} = ac^j \le  bc^{j + \mathbbm{1} [a \ge b]}$, and so subtour $Z_{j + \mathbbm{1}[a \ge b]}$ contains at least $l$ vertices. If the tour visits the $l$\textsuperscript{th} vertex before the subtour $Z_{j + \mathbbm{1}[a \ge b]}$, then
    \[
        T_l \le \sum_{m = 0}^{j - \mathbbm{1}[a < b]} 2 b c^m.
    \]
    Otherwise, let $u_l$ be the distance of the $l$\textsuperscript{th} vertex from the starting vertex on one side of cycle $C_{j + \mathbbm{1}[a \ge b])}$. Therefore, the distance from the other side is at most $2bc^{j + \mathbbm{1}[a \ge b])} - u_l$. Since the cycle is traversed from each direction with equal probability, in this case
    \[
        \E_b \big[T_l^p\big] \le \frac{1}{2} \bigg(\sum_{m = 0}^{j - \mathbbm{1}[a < b]} 2 b c^m + u_l \bigg) + \frac{1}{2} \bigg( \big(2b c^{j + \mathbbm{1}[a \ge b]} - u_l \big) + \sum_{m = 0}^{j - \mathbbm{1}[a < b]} 2 b c^m \bigg).
    \]
    
    For fixed $b, c$, this is maximum at $u_l = 0$ or $u_l = 2 b c^{j + \mathbbm{1}[a \ge b]}$. And therefore, we have that
    \begin{align*}
        \E_b \big[T_l^p\big] &\le \frac{1}{2} \bigg( \sum_{m = 0}^{j - \mathbbm{1}[a < b]} 2 b c^m \bigg)^p + \frac{1}{2} \bigg( \sum_{m = 0}^{j + \mathbbm{1}[a \ge b]} 2 b c^m\bigg)^p \\
        &= \frac{1}{2}\bigg( 2b \frac{c^{j + \mathbbm{1}[a \ge b]}-1}{c-1} \bigg)^p + \frac{1}{2}\bigg( 2b \frac{c^{j+1+ \mathbbm{1}[a \ge b]}-1}{c-1} \bigg)^p\\
        &\le \frac{1}{2}\bigg( 2b \frac{c^{j + \mathbbm{1}[a \ge b]}}{c-1} \bigg)^p + \frac{1}{2}\bigg( 2b \frac{c^{j+1+ \mathbbm{1}[a \ge b]}}{c-1} \bigg)^p\\
        &= \frac{2^{p-1}b^p (c^p + 1)}{(c-1)^p} c^{p(j + \mathbbm{1}[a \ge b])} \\
        &= c^{pj} \cdot \frac{2^{p-1}(c^p + 1)}{(c-1)^p} \big(b^pc^{p \cdot \mathbbm{1}[a \ge b]}\big).
    \end{align*}
        
    Taking expectation with respect to the random variable $b$ we have
    \begin{align*}
        \E \big[T_l^p\big] &\leq \E \bigg( {c^{pj} \cdot \frac{2^{p-1}(c^p + 1)}{(c-1)^p} \left(b^p c^{p \cdot \mathbbm{1}[a \ge b]}\right)} \bigg) \\
        &= c^{pj} \cdot \frac{2^{p - 1}(c^p + 1)}{(c-1)^p} \left( c^p\int_0^{\log_c{a}} c^{pU} dU  + \int_{\log_c{a}}^1 c^{pU} dU \right) \\
        &= c^{pj} \cdot \frac{2^{p - 1}(c^p + 1)}{(c-1)^p} \left( c^p \Big[ \frac{c^{pU}}{\ln c^p}\Big] \bigg|_{0}^{\log_c{a}}  + \Big[ \frac{c^{pU}}{\ln c^p}\Big] \bigg|_{\log_c{a}}^{1} \right) \\
        &= c^{pj} \frac{2^{p-1}(c^p + 1)}{(c-1)^p} \left( \frac{c^p(a^p-1)}{p \ln c} + \frac{c^p - a^p}{ p\ln c}\right) \\
        &= (ac^j)^p \cdot \frac{2^{p-1}(c^p + 1) (c^p-1)}{p(c-1)^p \ln c} \\
        &= \big(T_l^{OPT}\big)^p \cdot \frac{2^{p-1}(c^{2p} - 1)}{p(c-1)^p\ln c}.
    \end{align*}
    
    Let $f_p(c) = \frac{2^{p-1}(c^{2p} - 1)}{p(c-1)^p\ln c}$. This is the same function we minimize in \Cref{thm: lp_master_theorem_single_vehicle}, and we have that
    \[
        \min_{c \in (1, e)} f_p(c) \le \frac{8^p}{p \ln 4} \quad \forall\: p \ge 1,
    \]
    so that
    \[
        \E \big[T_l^p\big] \le \frac{8^p}{p \ln 4} \: \big(T_l^{OPT}\big)^p\quad \forall\: p \ge 1.
    \]
    This finishes the proof.
\end{proof}

\section{Multi Vehicle $L_p$~TSP}\label{sec:multi}

In this Section we develop a Linear Programming approach that provides state-of-the-art approximation guarantees for single/multi vehicle $L_p$~TSP problems.

\subsection{The Linear Programming Approach}\label{sec: lp}

We consider a natural linear programming relaxation of the $L_p$~TSP problem with $K$ vehicles, first introduced for a single vehicle TRP in \cite{CS11} and then generalized to multiple vehicles TRP in \cite{PS14}.
Before we begin, let us assume that pairwise distances, hence time, are always integral and polynomially bounded, with $d_e \in \{0\} \cup \big[O(n^2/\epsilon) \big]$ for each edge $e$, since quantization by rounding only adds a multiplicative error of $1 + O(\epsilon/n^2)$ to each edge and therefore results in an error of a factor at most $(1 + \epsilon)$ for the length of any path in the graph.

For all $t \ge 0$ and $i \in [K]$, let $\mathcal{P}_{t, i}$ (resp. $\mathcal{T}_{t, i}$) denote the set of all paths (resp. trees) of length at most $t$ rooted at vertex $s_i$. The linear program seeks to select paths from $\mathcal{P}_{t, i}$ at each time in $t \in \Z_+$ and cover the vertices progressively, stopping once each vertex has been covered. For all $v \in V$ and for all $t, i$, variable $x_{v, t, i}$ indicates if vertex $v$ has been visited exactly at time $t$ by vehicle $i$. For $t \in \Z_+$, variable $z_{P, t, i}$ indicates whether path $P \in \mathcal{P}_{t, i}$ has been selected at time $t$ (the maximum time $t$ can be upper bounded by $\sum_{e} d_e$, which as stated before can be assumed to be polynomially bounded). The objective function naturally is $\sum_{v, t, i} t^p x_{v, t, i}$. One natural constraint is that each vertex be visited at least once (constraint \eqref{eqn: lp-p-1}). Another natural constraint is that only one path from $\mathcal{P}_t$ must be selected at time $t$ (constraint \eqref{eqn: lp-p-2}). Finally, if vertex $v$ has been marked visited at some $t' \le t$, $v$ must lie on some path selected at time $t$ (constraint \eqref{eqn: lp-p-3}). We present the following linear programming relaxation using these constraints, denoted $LP_{\mathcal{P}, p, K}$, along with its dual $D_{\mathcal{P}, p, K}$. Let $OPT_{\mathcal{P}, p, K}$ denote its optimal value.

\begin{align}
    \min &\sum_{v, t, i} t^p x_{v, t, i} & \text{subject to}\\
    \sum_{t, i} x_{v, t, i} &\ge 1 & \forall \: v, \label{eqn: lp-p-1}\\
    \sum_{P \in \mathcal{P}_{t, i}} z_{P, t, i} &\le 1 & \forall \: t, i, \label{eqn: lp-p-2}\\
    \sum_{P \in \mathcal{P}_{t, i}: v \in P} z_{P, t, i} &\ge \sum_{t' \le t} x_{v, t', i} &\forall \: t, v, i, \label{eqn: lp-p-3} \\
    x, z &\ge 0 \notag.
\end{align}
\begin{align}
    \max &\sum_{v} \alpha_v - \sum_{t, i} \beta_{t, i} & \text{subject to} \\
    \alpha_v &\le t^p + \sum_{t' \ge t} \theta_{v, t', i} & \forall \: v, t, i, \label{eqn: d-p-1}\\
    \sum_{v \in P} \theta_{v, t, i} &\le \beta_{t, i} & \forall \: P \in \mathcal{P}_{t, i}, t, i, \label{eqn: d-p-2} \\
    \alpha, \beta, \theta &\ge 0 \notag.
\end{align}

\subsection{Solving The LP}\label{sec: lp_solution}

The above linear program has an exponential number of variables, and it is not known whether it can be solved directly. The next most natural line of attack -- solving the dual -- works here.

In the dual $D_{\mathcal{P}, p, K}$, separating over constraints \eqref{eqn: d-p-1} is straightforward since there are a polynomial number of variables and constraints involved. Constraints \eqref{eqn: d-p-2} have a combinatorial interpretation. Fix $i \in [K]$. Given rewards $\{ \theta_{v, t, i} \}$, is there some path rooted at $s_i$ of length at most $t$ (i.e., some path in $\mathcal{P}_{t, i}$) with total reward on the path more than $\beta_t$? This is the decision problem version of the \emph{path-orienteering problem}: given a starting vertex $s$, some nonnegative rewards for the vertices, and some budget $t > 0$, one needs to determine the maximum reward a path starting at $s$ of length at most $t$ can gather. 

Near-optimal approximations are known for the related \emph{tree-orienteering problem} (see \cite{CGRT03}). Consider the following relaxed version of the tree-orienteering problem: We are given nonnegative rewards for the vertices, a starting vertex $s$, and an approximation parameter $\epsilon > 0$. Let the reward gathered by the optimal path for the path-orienteering problem with length budget $t$ be $\beta^*$.  Then, is there some tree rooted at $s$ of length at most $(1 + \epsilon)t$ (i.e., some tree in $\mathcal{T}_{(1 + \epsilon)t, i}$) with total reward on the tree more than $\frac{\beta^*}{1 + \epsilon}$? \cite{CGRT03} design a polynomial time algorithm for determining gadgets called `good $l$-tees', and it also gives a polynomial-time algorithm for this version of the tree-orienteering problem using a reduction. To utilize this, we consider the following related LP (denoted $LP_{\mathcal{T}, p, K}$ and its dual, denoted $D_{\mathcal{T}, p, K}$):

\begin{align*}
    \min &\sum_{v, t, i} t^p x_{v, t, i} & \text{subject to} \\
    \sum_{t, i} x_{v, t, i} &\ge 1 & \forall \: v,\\
    \sum_{T \in \mathcal{T}_{t, i}} z_{T, t, i} &\le 1 & \forall \: t, i,\\
    \sum_{T \in \mathcal{T}_{t, i}: v \in T} z_{T, t, i} &\ge \sum_{t' \le t} x_{v, t', i} &\forall \: t, v, i, \\
    x, z &\ge 0.
\end{align*}
\begin{align*}
    \max &\sum_{v} \alpha_v - \sum_{t, i} \beta_{t, i} &\text{subject to} \\
    \alpha_v &\le t^p + \sum_{t' \ge t} \theta_{v, t', i} & \forall \: v, t, i, \\
    \sum_{v \in T} \theta_{v, t, i} &\le \beta_{t, i} & \forall \: T \in \mathcal{T}_{t, i}, t, i, \\
    \alpha, \beta, \theta &\ge 0. 
\end{align*}

We still cannot separate the dual for this linear program using tree-orienteering. To utilize tree-orienteering, we need to introduce the parameter $\eta = 1 + \epsilon$ (where $\epsilon > 0$) in the linear program. We denote this modification of $LP_{\mathcal{T}, p, K}$ as $LP^\eta_{\mathcal{T}, p, K}$, and its dual is denoted $D^\eta_{\mathcal{T}, p, K}$.

\begin{align}
    \min &\sum_{v, t, i} t^p x_{v, t, i} &\text{subject to } \notag\\
    \sum_{t, i} x_{v, t, i} &\ge 1 & \forall \: v, \label{eqn: lp-t-approx-1}\\
    \sum_{T \in \mathcal{T}_{\eta t, i}} z_{T, t, i} &\le \eta & \forall \: t, i, \label{eqn: lp-t-approx-2}\\
    \sum_{T \in \mathcal{T}_{\eta t, i}: v \in T} z_{T, t, i} &\ge \sum_{t' \le t} x_{v, t', i} &\forall \: t, v, i, \label{eqn: lp-t-approx-3} \\
    x, z &\ge 0. \notag 
\end{align}
\begin{align}
    \max & \sum_{v} \alpha_v - \eta \sum_{t, i} \beta_{t, i}  & \text{subject to} \notag \\
    \alpha_v &\le t^p + \sum_{t' \ge t} \theta_{v, t', i} & \forall \: v, t, i, \label{eqn: d-t-approx-1} \\
    \sum_{v \in T} \theta_{v, t, i} &\le \beta_{t, i} & \forall \: T \in \mathcal{T}_{\eta t, i}, t, i, \label{eqn: d-t-approx-2} \\
    \alpha, \beta, \theta &\ge 0. \label{eqn: d-t-approx-3}.
\end{align}

Post and Swami \cite{PS14} claim that for any $\epsilon > 0$, a feasible solution to $LP^\eta_{\mathcal{T}, 1, K}$ of cost at most $OPT_{\mathcal{P}, 1, K}$ can be computed in polynomial time, using arguments analogous to those in Lemma 3.2 in \cite{CS11}. We generalize their proof to arbitrary $p > 0$; the proofs are similar, but we include it here for completeness.

\begin{lemma}\label{lem: solvability}
    Given any $\epsilon > 0$ and $p > 0$, a feasible solution to $LP^{1 + \epsilon}_{\mathcal{T}, p, K}$ can be found with cost at most $OPT_{\mathcal{P}, p, K}$.
\end{lemma}

\begin{proof}
    The main idea of the proof is to provide a modified separation oracle for the Ellipsoid algorithm. Recall that to solve linear programs over a polytope $Q \in \R^m$ in polynomial time, the ellipsoid algorithm requires a polynomial time separation oracle that does the following: when a point $x_0 \in \R^m$ is provided to the oracle, the oracle output either that (1) $x_0 \in Q$, or (2) gives a separating hyperplane $d^T x \le \delta$ such that $d^T x_0 > \delta$ but $d^T x \le \delta$ for all $x \in Q$, (i.e., $x_0$ and $Q$ lie on opposite sides of this hyperplane). Given this oracle, the ellipsoid algorithm in polynomial time either (1) outputs that $Q = \emptyset$, or (2) gives a feasible point in $Q$.
    
    Now suppose we have another polytope $Q'$ such that $Q \subseteq Q'$. For any objective $c$, $\min_{x \in Q'} c^T x \le \min_{x \in Q}$. Let $f: \R^m \to \R^m$ be an arbitrary function. Suppose we have a polynomial-time separation oracle that given a point $x_0$, the oracle outputs either that (1) $f(x_0) \in Q'$, or (2) gives a hyperplane separating $x_0$ and $Q$. Then, the ellipsoid algorithm in polynomial time either (1) outputs that $Q = \emptyset$, or (2) gives a feasible point $x'$ in $Q'$. 

    Let us first consider the relevant polytopes:
    \begin{align*}
        P_{\mathcal{P}, \nu, \eta} &= \Big\{(\alpha, \beta, \theta): (\alpha, \beta, \theta) \:\text{satisfy}\: \eqref{eqn: d-t-approx-1}, \eqref{eqn: d-t-approx-3}, \sum_{v \in P} \theta_{v, t, i} \le \beta_{t, i} \: \forall \: P \in \mathcal{P}_{\eta t, i}, t, i, \:\text{and}\\
        & \qquad \sum_{v} \alpha_v - \eta \sum_{t, i} \beta_{t, i} \ge \nu \Big\} \\
        P_{\mathcal{T}, \nu, \eta} &= \Big\{(\alpha, \beta, \theta): (\alpha, \beta, \theta) \:\text{satisfy}\: (\eqref{eqn: d-t-approx-1}), (\eqref{eqn: d-t-approx-2}), (\eqref{eqn: d-t-approx-3}), \:\text{and}\: \sum_{v} \alpha_v - \eta \sum_{t, i} \beta_{t, i} \ge \nu \Big\}.
    \end{align*}
    
    We give an oracle that given a $\nu \in \R$ and $(\alpha, \beta, \theta)$ outputs either (1) a confirmation that $(\alpha, \eta \beta, \theta) \in P_{\mathcal{P}, \nu, 1}$ or (2) a separating hyperplane for point $(\alpha, \beta, \theta)$ and polytope $P_{\mathcal{T}, \nu, \eta}$. Since $P_{\mathcal{T}, \nu, \eta} \subseteq P_{\mathcal{P}, \nu, 1}$, the ellipsoid algorithm in polynomial time gives either (1) a confirmation that $P_{\mathcal{T}, \nu, \eta} = \emptyset$, or (2) a point in $P_{\mathcal{P}, \nu, 1}$. 
    
    Using binary search, find the largest $\nu^*$ such that the ellipsoid algorithm gives a point in $P_{\mathcal{P}, \nu, 1}$. Since $OPT_{\mathcal{P}, p, K}$ is the largest value of $\nu$ such that $P_{\mathcal{P}, \nu, 1}$ is non-empty, this implies that $\nu^* \le OPT_{\mathcal{P}, \nu, 1}$. For each $\delta > 0$, the ellipsoid algorithm gives a confirmation that $P_{\mathcal{T}, \nu^* + \delta, \eta} = \emptyset$ by giving an inconsistent subsystem of inequalities for $P_{\mathcal{T}, \nu^* + \delta, \eta}$. This subsystem has all inequalities from constraints \eqref{eqn: d-t-approx-1}, \eqref{eqn: d-t-approx-3}, and the inequality $\sum_{v} \alpha_v - \eta \sum_{t, i} \beta_{t, i} \ge \nu^* + \delta$. It also contains only a polynomial number of inequalities from \eqref{eqn: d-t-approx-2}. By complementary slackness, we can get a solution to $LP_{\mathcal{T}, P, K}$ of cost at most $\nu^* + \delta$ with a polynomial number of nonzero variables. Choose $\delta$ to be small enough so that this cost is at most $\nu^* \le OPT_{\mathcal{P}, \nu, 1}$ but large enough so that the algorithm is still polynomial time.
    
    We describe the separation oracle now. Checking whether $\sum_{v} \alpha_v - \sum_{t, i} \eta \beta_{t, i} \ge \nu$ is easy, and if it is not true, $\sum_{v} \alpha_v - \sum_{t, i} \beta_{t, i} \ge \nu$ can be output as a separating hyperplane between $(\alpha, \beta, \theta)$ and $P_{\mathcal{T}, \nu, \eta}$. Similarly, each inequality in \eqref{eqn: d-t-approx-1}, \eqref{eqn: d-t-approx-3} for $P_{\mathcal{P}, \nu, 1}$ can be checked individually and if $(\alpha, \eta \beta, \theta)$ does not satisfy any of those, that inequality can be output as a separating hyperplane between $(\alpha, \beta, \theta)$ and $P_{\mathcal{T}, \nu, \eta}$. Since there are a polynomial number of these inequalities, this can be done in polynomial time.
    
    It remains to check inequalities \eqref{eqn: d-t-approx-2} for $P_{\mathcal{P}, \nu, 1}$. To this end, for each $t, i$, solve the relaxed version of the tree-orienteering problem with starting vertex $s_i$, rewards $\{\theta_{v, t, i}: v \in V \}$, approximation parameter $\epsilon$, and length budget $t$ to get a tree $T \in \mathcal{T}_{\eta t, i}$. $T$ has reward at least $1/\eta$ times the optimal reward for path-orienteering with length bound $t$. If for some $t, i$, the reward for $T \in \mathcal{T}_{\eta t, i}$ exceeds $\beta_{t, i}$, then $\sum_{v \in T} \theta_{v, t, i} \le \beta_{t, i}$ as the separating hyperplane between $(\alpha, \beta, \theta)$ and $P_{\mathcal{T}, \nu, \eta}$. Otherwise, all paths in $\mathcal{P}_{t, i}$ gather reward at most $\eta \beta_{t, i}$, and therefore $(\alpha, \eta \beta, \theta) \in P_{\mathcal{P}, \nu, 1}$. This proves the correctness of the separation oracle and therefore finishes the proof.
\end{proof}

\subsection{The Rounding Algorithm} \label{sec: rounding_algorithm}

Using \Cref{lem: solvability}, we generalize the LP-rounding algorithm of Post and Swami \cite{PS14} for TRP (i.e., $L_1$~TSP) to $L_p$~TSP, achieving approximation algorithms for the $L_p$~TSP problem in general and TFP (i.e., $L_2$-TSP) in particular, for both the single-vehicle and the multi-vehicle settings.
 
Let $\ell^{OPT} = (\ell^{OPT}_v: v \in V)$ denote the optimal visit times for a given instance of the $L_p$~TSP problem with $K$ vehicles starting at vertices $s_1, \ldots, s_K$ respectively, with the corresponding cost $\|\ell^{OPT}\|_p = \Big( \sum_{v \in V} \big(\ell_v^{OPT}\big)^p\Big)^{1/p}$. Since $LP_{\mathcal{P}, p, K}$ is a relaxation for this problem, we have that the optimal cost for it, $OPT_{\mathcal{P}, p, K}$ is at most $\|\ell^{OPT}\|_p^p$. For any $\epsilon > 0$ (that we fix later), using \Cref{lem: solvability}, obtain a solution $(x^*, z^*)$ to $LP^{1 + \epsilon}_{\mathcal{T}, p, K}$ with cost at most $OPT_{\mathcal{P}, p, K} \le \|\ell^{OPT}\|_p^p$. For convenience, denote $\ell_v^{LP} = \big(\sum_{t, i} t^p x^*_{v, t, i} \big)^{1/p}$, and $\ell^{LP} = \Big( \sum_{v \in V} \big(\ell_v^{LP}\big)^p\Big)^{1/p} = \big(OPT_{\mathcal{P}, p, K}\big)^{1/p}$. Therefore, we have that $\|\ell^{OPT}\|_p \ge \ell^{LP}$.
 
We give a randomized rounding algorithm (called LP-ROUND) with cost $\|\ell^R\|_p = \big( \sum_{v \in V} (\ell_v^R)^p \big)^{1/p}$. The algorithm takes as input an the solution $(x^*, z^*)$ to $LP^\eta_{\mathcal{T}, p, K}$. The fundamental idea for rounding is the same as that in \cite{CGRT03} and \cite{FTT21}: get exponentially large tours using trees with lengths bounded in terms of lengths of paths, except that instead of choosing good $l$-trees, we choose trees using the solution $(x^*, z^*)$. The major advantage this presents is that it is relatively easy to deal with the multi-vehicle case.

\begin{algorithm}[H]
\caption{Randomized LP Rounding for Routing}\label{alg: lp-round}
\begin{algorithmic}[1]
    \State Let $c \in (1, e)$ be a constant that we fix later, and $b = c^U$, where $U \sim U[0, 1)$, that is, $U$ is sampled uniformly from $[0, 1)$. Denote $t_j = bc^j$ for all $j \ge 0$
    
    \For {$j = 0, 1, \ldots $,}:
        \For{$i \in [K]$}:
            \State - Choose a tree $T_{i, j}$ from the distribution $\big\{ z^*_{T, t_j, i}/\eta: T \in \mathcal{T}_{\eta t_j, i}\big\}$ independently
            \State - Double $T_{i, j}$ to get a cycle $C_{i, j}$, short-cutting any vertices already visited
            \State - Traverse $C_{i, j}$ from $s_i$ in either direction with probability $1/2$ to get a tour $Z_{i, j}$
        \EndFor
        
        \State Exit if every vertex has been covered by some tour
    \EndFor
    
    \For{$i \in [K]$}
        \State Concatenate the subtours $Z_{i, 0}, Z_{i, 1}, \ldots $ to obtain the tour $Z_i$ for vehicle $i$
    \EndFor
    \State \Return the first visit times of vertices in $V$ among tours $\{Z_i: i \in [K]\}$
\end{algorithmic}
\end{algorithm}

 The analysis of the algorithm is somewhat involved and we introduce some notation to help. Let $p_{v, j}$ be the probability that vertex $v$ is not covered by the end of iteration $j$. For convenience, define $t_{-1} = 0$ and $p_{v, -2} = 0$ for all $v$. Observe that the probability that vertex $v$ is covered in iteration $j$ is $p_{v, j - 1} - p_{v, j}$. Also denote $y_{v, t, i} = \sum_{t' \le t} x^*_{v, t', i}$ for all $v, t, i$, and $w_{v, j} = 1 - \sum_{i} y_{v, t_j, i}$. When $K = 1$, we will omit the index $i$ corresponding to the vehicle in all variables.

We first need a recurrence relation for the probabilities $p_{v, j}$.

\begin{lemma}\label{lemma: probability_bound}
    \begin{enumerate}
        \item When $K = 1$, for arbitrary $s_1$,
        \[
            p_{v, j} \le \frac{1}{\eta} w_{v, j} + \big(1 - \frac{1}{\eta}\big) p_{v, j - 1} \qquad \forall\: v \in V, j \ge -1.
        \]
        \item For an arbitrary $K$ and arbitrary $s_1, \ldots, s_K$,
        \[
            p_{v, j} \le \big(1 - e^{-1/\eta}\big) w_{v, j} + e^{-1/\eta} p_{v, j - 1} \qquad \forall\: v \in V, j \ge -1.
        \]
    \end{enumerate} 
\end{lemma}

\begin{proof}
    \begin{enumerate}
        \item The probability that $v$ is visited in iteration $j$ is $\sum_{T \in \mathcal{T}_{\eta t_j}: v \in T} \frac{z_{T, t_j}}{\eta} \ge \frac{y_{v, t_j}}{\eta}$. Therefore,
        \[
            p_{v, j} \le p_{v, j - 1}\Big(1 - \frac{y_{v, t}}{\eta}\Big) = p_{v, j - 1} \Big(1 - \frac{1 - w_{v, j}}{\eta} \Big) = \Big(1 - \frac{1}{\eta}\Big) p_{v, j -1} + \frac{1}{\eta} w_{v, j}. 
        \]
        
        \item This is Lemma 5.3 of \cite{PS14} and follows an argument similar to part 1. For completeness, we include the proof here. The probability that $v$ is visited by vehicle $i$ in iteration $j$ is
        \[
            \sum_{T \in \mathcal{T}_{\eta t_j, i}: v \in T} \frac{z_{T, t_j}}{\eta} \ge \frac{y_{v, t_j}}{\eta}.
        \]
        Therefore, $p_{v, j} \le p_{v, j - 1} \prod_i \big(1 - \frac{y_{v, t_j, i}}{\eta}\big)$. Now,
        \begin{align*}
            \prod_i \Big(1 - \frac{y_{v, t_j, i}}{\eta}\Big) &\le \Big(1 - \frac{\sum_{i} y_{v, t_j, i}}{\eta K}\Big)^K \\
            &= \Big(1 - \frac{1 - w_{v, j}}{\eta K}\Big)^K \\
            &\le \Big(1 - \frac{1}{\eta K}\Big)^K + \Big(1 - \big(1 - \frac{1}{\eta K} \big)^K\Big) w_{v, j} \\
            &\le e^{-1/\eta} + (1 - e^{-1/\eta})w_{v, j}.
        \end{align*}
        
        The first inequality is the AM-GM inequality, the second follows by convexity of $f(x) = \big(1 - \frac{1 - x}{\eta K}\big)^K$, and the third is a standard bound for the exponential function. The result follows since $p_{v, j - 1} \le 1$ for all $j \ge -1$.
    \end{enumerate}
\end{proof}

We define some additional terms that will help us bound $\ell^R_v$ in terms of $\ell^{LP}_v$ for each vertex $v$. Define $\Delta_j = (bc^j)^p - (bc^{j - 1})^p = t_j^p - t_{j - 1}^p$ for all $j \ge 0$. Denote 
\[
    \beta_v = \sum_{j \ge 0} p_{v, j - 1} \big( (bc^j)^p - (bc^{j - 1})^p\big) = \sum_{j \ge 0} p_{v, j - 1} \Delta_j, \qquad \alpha_v = \sum_{j \ge 0} w_{v, j - 1} \Delta_j.
\]
For all reals $t \ge 0$, let $\sigma(t)$ be the smallest $t_j^p$ such that $t_j \ge t$, that is, $\sigma(t) = \min_{t_j \ge t} t_j^p$. For a random variable $X$, let $\E_b X$ denote the conditional expectation of $X$ for a fixed $b$.

\begin{lemma}\label{lem: latencies-relations} For each vertex $v \in V$,
    \begin{enumerate}
        \item $\E \alpha_v = \frac{c^p - 1}{p \ln c} \big(\ell_v^{LP}\big)^p$,
        \item $\E_b \big[ \big(\ell_v^R\big)^p \big] \le \frac{2^{p - 1} (c^p + 1) \eta^p}{(c - 1)^p} \beta_v$.
    \end{enumerate}
\end{lemma}

\begin{proof}
    \begin{enumerate}
        \item We first show that $\alpha_v = \sum_{t, i} \sigma(t) x^*_{v, t, i}$.
    \begin{align*}
        \sum_{t, i} \sigma(t) x^*_{v, t, i} &= \sum_{j \ge 0} t_j^p \sum_{t = t_{j - 1} + 1}^{t_j} \sum_i x^*_{v, t, i} \\
        &= \sum_{j \ge 0} \sum_{j' = 0}^j \Delta_{j'} \sum_{t = t_{j - 1} + 1}^{t_j} \sum_i x^*_{v, t, i} \\
        &= \sum_{j' \ge 0} \Delta_{j'} \sum_{j \ge j'} \sum_{t = t_{j - 1} + 1}^{t_j} \sum_i x^*_{v, t, i} \\
        &= \sum_{j' \ge 0} \Delta_{j'} w_{v, j' - 1} = \alpha_v
    \end{align*}

    This helps determine $\E \alpha_v$ in terms of $\ell_v^{LP}$. Suppose $t = ac^j$ for some real $a \in [1, c)$ and integer $j \ge 0$. Then, $\sigma(t) = \big(b c^{j + \mathbbm{1}[a > b]}\big)^p$. Therefore, since $b = c^U$ where $U \sim U[0, 1)$,
    \begin{align*}
        \E\big(\sigma(t)\big) &= \int_0^{a} c^{pU} c^{p(j + 1)} dU + \int_{a}^{1} c^{pU} c^{pj} dU = \frac{c^p - 1}{p \ln c} \cdot t^p.
    \end{align*}
    
    Consequently, since $\alpha_v = \sum_{t, i} \sigma(t) x^*_{v, t, i}$, we have that
    \begin{equation}\label{eqn: relate_latencies_1}
        \E \alpha_v = \frac{c^p - 1}{p \ln c} \cdot\big(\ell_v^{LP}\big)^p.
    \end{equation}
        
        \item Suppose $v$ is covered in the $j$th subtour $Z_{i, j}$ for some vehicle $i$. The cycle $C_{i, j}$ in iteration $j$ has length at most $2 \eta t_j = 2 \eta b c^j$. Let $u_v$ be the distance of vertex $v$ from the starting vertex in one direction of the cycle $C_{i, j}$. Then, the distance of $v$ from the starting vertex in the other direction of $C_{i, j}$ is at most $2 \eta b c^j - u_v$. Since each direction is taken with probability $1/2$, we have in this case that
    
        \[
            \big(\ell_v^R\big)^p \le \frac{1}{2} \big(\eta (2 b c^0 + \ldots + 2 b c^{j - 1} + u_v)\big)^p + \frac{1}{2} \big(\eta (2 b c^0 + \ldots + 2 b c^{j - 1} + (2 b c^j - u_v))\big)^p.
        \]
        For fixed $b, c$, the right hand side is maximum at $u_v = 0$ or $u_v = 2 b c^j$, so that
        
        \begin{align*}
            \big(\ell_v^R\big)^p &\le \frac{1}{2} \big(\eta (2 b c^0 + \ldots + 2 b c^{j - 1} + d_v)\big)^p + \frac{1}{2} \big(\eta (2 b c^0 + \ldots + 2 b c^{j - 1} + (2 b c^j - d_v))\big)^p.
            \\
            &\le \frac{1}{2} \big(\eta (2 b c^0 + \ldots + 2 b c^{j - 1})\big)^p + \frac{1}{2} \big(\eta (2 b c^0 + \ldots + 2 b c^{j - 1} + 2 b c^j )\big)^p \\
            &\le \frac{(2 \eta b c^j)^p}{2}\Big[\Big(\frac{1}{c - 1}\Big)^p + \Big(\frac{c}{c - 1}\Big)^p \Big] \\
            &= \frac{2^{p - 1} (c^p + 1)}{(c - 1)^p} \cdot (\eta b c^j)^p.
        \end{align*}
        
        Since the probability of vertex $v$ being covered in the $j$th iteration is $p_{v, j} - p_{v, j - 1}$, the above implies that
        \begin{align*}
            \E_b \Big[ \big(\ell_v^R\big)^p \Big] &\le \sum_{j \ge 0} (p_{v, j } - p_{v, j - 1})\cdot \frac{2^{p - 1} (c^p + 1)}{(c - 1)^p} \cdot (\eta b c^j)^p \\
            &= \frac{2^{p - 1} (c^p + 1) \eta^p}{(c - 1)^p} \sum_{j \ge 0} p_{v, j - 1} \big( (bc^j)^p - (bc^{j - 1})^p\big) \\
            &= \frac{2^{p - 1} (c^p + 1) \eta^p}{(c - 1)^p} \beta_v.
        \end{align*}
    \end{enumerate}
\end{proof}

We are ready to present our main results. To warm up, let us first analyze the approximation bound for single vehicle case, providing matching guarantees to that of the combinatorial approach.

\begin{theorem}\label{thm: lp_master_theorem_single_vehicle}
Given optimal solutions to $LP^\eta_{\mathcal{T}, p, K}$, randomized rounding
    \Cref{alg: lp-round} provides solutions that are
    \begin{enumerate}
        \item $4.27$-approximation for single-vehicle TFP,
        \item $\frac{8}{(p \ln 4)^{1/p}}$-approximation for single-vehicle $L_p$~TSP for any $p \ge 1$.
    \end{enumerate}
\end{theorem}

\begin{proof}
    We will show that for all $v \in V$,
    \begin{equation} \label{eq: lp-vertex-bound}
        \E\big[\big(\ell_v^R\big)^p\big] \le C \big(\ell^{LP}_v\big)^p
    \end{equation}
    for an appropriate constant $C$. As observed earlier, this implies that
    \[
        \E \big[ \|\ell^R\|_p^p \big] \le C (\ell^{LP})^p \le C \|\ell^{OPT}\|_p^p.
    \]
    By Markov's inequality, for any constant $\tau > 0$,
    \[
        \Pr\Big(\|\ell^R\|_p^p > (1 + \tau)\E \big[ \|\ell^R\|_p^p \big] \Big) \le \frac{1}{1 + \tau}.
    \]
    
    For $\tau \in (0, 1)$, after $m = \frac{2 \log n}{\tau}$ independent runs of the rounding algorithm, the probability that $\|\ell^R\|_p^p > (1 + \tau)\E \big[ \|\ell^R\|_p^p \big]$ in each run of the experiment is at most
    \[
        \Big(\frac{1}{1 + \tau}\Big)^m \le \big(e^{-\tau/2}\big)^m = \frac{1}{n}.
    \]
    
    Therefore, with high probability, after $2 \log{n}/\tau$ independent runs of the algorithm, there will be at least one run of the algorithm where
    \[
        \|\ell^R\|_p \le (1 + \tau)^{1/p} \big(\E \big[ \|\ell^R\|_p^p \big] \big)^{1/p} \le (1 + \tau)^{1/p} \cdot C^{1/p} \|\ell^{OPT}\|_p \le (1 + \tau)C^{1/p} \cdot \|\ell^{OPT}\|_p,
    \]
    
    so that enough independent runs of LP-ROUND give a $\big((1 + \tau) C^{1/p}\big)$-approximation algorithm for $L_p$~TSP for all constants $\tau \in (0, 1)$. This in particular implies that LP-ROUND is a $C_0^{1/p}$-approximation for any $C_0 > C$, using an appropriate constant $\tau$ satisfying $(1 + \tau)C^{1/p} \le C_0^{1/p}$.
    
    We proceed to prove inequality \eqref{eq: lp-vertex-bound} and specify the constant $C$ using Lemmas \ref{lemma: probability_bound} and \ref{lem: latencies-relations}.
    
    \medskip
    By \Cref{lemma: probability_bound} part (1) and the fact that $p_{v, -2} = 0$, we have
    \begin{align*}
        \beta_v &\le \frac{1}{\eta} \sum_{j \ge 0} w_{v, j - 1} \Delta_j + \Big(1 - \frac{1}{\eta}\Big) \sum_{j \ge 0} p_{v, j - 2} \Delta_{j} \\
        &= \frac{\alpha_v}{\eta}  + \Big(1 - \frac{1}{\eta}\Big) \sum_{j \ge 0} p_{v, j - 1} \Delta_{j + 1} \\
        &= \frac{\alpha_v}{\eta} + c^p \Big(1 - \frac{1}{\eta}\Big) \sum_{j \ge 0} p_{v, j - 1} \Delta_j \\
        &= \frac{\alpha_v}{\eta} + c^p \Big(1 - \frac{1}{\eta}\Big) \beta_v.
    \end{align*}
    
    This implies that
    \[
        \beta_v \le \frac{\alpha_v}{\eta\bigg(1 - \Big(1 - \frac{1}{\eta}\Big) c^p \bigg)},
    \]
    and consequently using \Cref{lem: latencies-relations} part (2) that
    \[
        \E_b \Big[ \big(\ell_v^R\big)^p \Big] \le \frac{2^{p - 1} (c^p + 1)}{(c - 1)^p} \frac{\eta^p}{\eta\bigg(1 - \Big(1 - \frac{1}{\eta}\Big) c^p \bigg)} \cdot \alpha_v.
    \]
    
    Taking expectation in $b$ on both sides and using \Cref{lem: latencies-relations} part (1), we get that
    \begin{align*}
        \E \Big[ \big(\ell_v^R\big)^p \Big] &\le \frac{2^{p - 1} (c^p + 1) \eta^p}{(c - 1)^p} \cdot \frac{\eta^p}{\eta\bigg(1 - \Big(1 - \frac{1}{\eta}\Big) c^p \bigg)} \cdot \E \alpha_v \\ &= \frac{2^{p - 1} (c^{2p} - 1) \eta^p}{p (c - 1)^p \ln c} \cdot \frac{\eta^p}{\eta\bigg(1 - \Big(1 - \frac{1}{\eta}\Big) c^p \bigg)} \cdot \big(\ell_v^{LP}\big)^p.
    \end{align*}
    
    Since
    \[
        \lim_{\epsilon \to 0^+} \frac{\eta^p}{\eta\bigg(1 - \Big(1 - \frac{1}{\eta}\Big) c^p \bigg)} = \lim_{\eta \to 1^+} \frac{\eta^p}{\eta\bigg(1 - \Big(1 - \frac{1}{\eta}\Big) c^p \bigg)} = 1,
    \]
    for each $\delta > 0$, there exists a $\eta - 1 = \epsilon > 0$ such that
    \[
       \frac{2^{p - 1} (c^{2p} - 1)}{p (c - 1)^p \ln c} \cdot \frac{\eta^p}{\eta\bigg(1 - \Big(1 - \frac{1}{\eta}\Big) c^p \bigg)} \le  \frac{2^{p - 1} (c^{2p} - 1)}{p (c - 1)^p \ln c} + \delta.
    \]
    
    Therefore, for any given $\delta > 0$, we have some $\epsilon > 0$ so that
    \begin{equation*}
        \E \Big[ (\ell_v^R)^p \Big] \le \bigg( \frac{2^{p - 1} (c^{2p} - 1)}{p (c - 1)^p \ln c} + \delta \bigg) \big(\ell_V^{LP}\big)^p.
    \end{equation*}
    
    That is, the constant $C$ in inequality \eqref{eq: lp-vertex-bound} equals $\min_{c \in (1, e)} \frac{2^{p - 1} (c^{2p} - 1)}{p (c - 1)^p \ln c} + \delta$.
    
    \begin{enumerate}
        \item $p = 2$. Given any $\delta' > 0$, choose $c = e - \epsilon'$ for a small enough $\epsilon'$ so that $\frac{c^4 - 1}{(c - 1)^2 \ln c} \le \frac{e^4 - 1}{(e - 1)^2} + \delta' \le 18.154 + \delta'$. Finally, choose a small enough $\delta$ so that $18.154 + \delta' + \delta < (4.27)^2$. This gives that ROUND is a $4.27$-approximation to single-vehicle TFP.
    
        \item Arbitrary $p \ge 1$. Using an argument similar to part 1, it is enough to show that $\min_{c \in (1, e)} \Big(\frac{2^{p - 1} (c^{2p} - 1)}{p (c - 1)^p \ln c}\Big)^{1/p} < \frac{8}{(p \ln 4)^{1/p}}$ for all $p \ge 1$. We have
        \[
            \Big(\frac{2^{p - 1} (c^{2p} - 1)}{p (c - 1)^p \ln c}\Big)^{1/p} \le \Big(\frac{2^{p - 1} c^{2p}}{p (c - 1)^p \ln c}\Big)^{1/p} = \frac{2c^2}{c - 1} \Big(\frac{1}{2 p \ln c}\Big)^{1/p}.
        \]
        Choosing $c = 2$ gives
        \[
           \min_{c \in (1, e)} \Big(\frac{2^{p - 1} (c^{2p} - 1)}{p (c - 1)^p \ln c}\Big)^{1/p} \le \frac{8}{\big(p \ln 4 \big)^{1/p}}.
        \]
        
        It is easy to see that the minimum is not achieved at $c = 2$, so that the above inequality is strict, proving the claim.
    \end{enumerate}
\end{proof}

We end this section by proving the promised bounds for multi-vehicle problems.

\begin{theorem}\label{thm: lp_master_theorem_multi_vehicle}
    Given optimal solutions to $LP^\eta_{\mathcal{T}, p, K}$, the randomized rounding \Cref{alg: lp-round} provides solutions that are
    \begin{enumerate}
        \item $10.92$-approximation for multi-vehicle TFP,
        \item $18 p$-approximation for multi-vehicle $L_p$~TSP for all $p \ge 1$.
    \end{enumerate}
\end{theorem}

\begin{proof}
    Following the argument in \Cref{thm: lp_master_theorem_single_vehicle}, it is enough to show that for all $v \in V$, $\E\big[\big(\ell_v^R\big)^p\big] \le C \big(\ell^{LP}_v\big)^p$ for an appropriate $C$.
     
    By \Cref{lemma: probability_bound} part (2) and the fact that $p_{v, -2} = 0$, we have that
    \begin{align*}
        \beta_v &= \sum_{j \ge 0} p_{v, j - 1} \Delta_j \\
        &\le (1 - e^{-1/\eta}) \sum_{j \ge 0} w_{v, j - 1} \Delta_j + e^{-1/\eta} \sum_{j \ge 0} p_{v, j - 2} \Delta_{j} \\ &= (1 - e^{-1/\eta}) \alpha_v  + e^{-1/\eta} \sum_{j \ge 0} p_{v, j - 1} \Delta_{j + 1} \\
        &= (1 - e^{-1/\eta}) \alpha_v + e^{-1/\eta} c^p \sum_{j \ge 0} p_{v, j - 1} \Delta_j \\
        &= (1 - e^{-1/\eta}) \alpha_v + e^{-1/\eta} c^p \beta_v.
    \end{align*}
    
    This implies that
    \[
        \beta_v \le \frac{1 - e^{-1/\eta}}{1 - c^p e^{-1/\eta}} \cdot \alpha_v
    \]
    and consequently using \Cref{lem: latencies-relations} that
    \[
        \E \Big[\big(\ell_v^R\big)^p\Big] \le \frac{2^{p - 1} (c^{p} + 1)}{(c - 1)^p} \cdot \frac{\eta^p (1 - e^{-1/\eta})}{1 - c^p e^{-1/\eta}} \cdot \E \alpha_v = \frac{2^{p - 1} (c^{2p} - 1)}{p (c - 1)^p \ln c} \cdot \frac{\eta^p (1 - e^{-1/\eta})}{1 - c^p e^{-1/\eta}} \cdot \big(\ell_v^{LP}\big)^p
    \]
    Since
    \[
        \lim_{\epsilon \to 0^+} \frac{\eta^p (1 - e^{-1/\eta})}{1 - c^p e^{-1/\eta}} = \lim_{\eta \to 1^+} \frac{\eta^p (1 - e^{-1/\eta})}{1 - c^p e^{-1/\eta}} = \frac{e - 1}{e - c^p},
    \]
    for each $\delta > 0$, there exists a $\eta - 1 = \epsilon > 0$ such that
    \[
        \frac{\E \Big[ \big(\ell_v^R\big)^p\Big]}{\big(\ell_v^{LP}\big)^p} \le \frac{2^{p - 1} (c^{2p} - 1)}{p (c - 1)^p \ln c} \cdot \frac{\eta^p (1 - e^{-1/\eta})}{1 - c^p e^{-1/\eta}} \le \frac{(e - 1)2^{p - 1}(c^{2p} - 1)}{p (c - 1)^p (e  - c^p) \ln c} + \delta.
    \]
    
    Therefore, the constant $C$ in inequality \eqref{eq: lp-vertex-bound} equals $\min_{c \in (1, e^{1/p})}\frac{(e - 1)2^{p - 1}(c^{2p} - 1)}{p (c - 1)^p (e  - c^p) \ln c} + \delta$.
    
    \begin{enumerate}
        \item  $p = 2$. Choose $c = 1.834$, so that $\min_{c \in (1, e^{1/p})}\frac{(e - 1)(c^4 - 1)}{2(c - 1)^2(e - c^2) \ln c} \le 119.21$. Choose $\delta$ small enough so that $119.21 + \delta < (10.92)^2$, giving that ROUND is a $10.92$-approximation for multi-vehicle TSP.
        
        \item Arbitrary $p \ge 1$. Let $f_p(c) = \Big(\frac{(e - 1)2^{p - 1}(c^{2p} - 1)}{p (c - 1)^p (e  - c^p) \ln c}  
        \Big)^{1/p} = \frac{2}{c - 1} \Big(\frac{(e - 1)(c^{2p} - 1)}{2(e  - c^p) \ln c^p}
        \Big)^{1/p}$. Using an argument similar to part 1, it is enough to show that $\min_{c \in (1, e^{1/p})} < 18 p$.
        
        Let $\theta = c^p$; we choose $\theta$ later. Then using $e^x \ge 1 + x$, we have
        \begin{align*}
            f_p(\theta^{1/p}) &= \frac{2}{\theta^{1/p} - 1} \bigg(\frac{(e - 1)(\theta^2 - 1)}{2(e  - \theta) \ln \theta} \bigg)^{1/p} \\
            &= \frac{2}{e^{\ln \theta/p} - 1} \bigg(\frac{(e - 1)(\theta^2 - 1)}{2(e  - \theta) \ln \theta} \bigg)^{1/p} \\
            &\le \frac{2p}{\ln \theta} \bigg(\frac{(e - 1)(\theta^2 - 1)}{2(e  - \theta) \ln \theta} \bigg)^{1/p}.
        \end{align*}
        
        Choose $\theta = 1.152$, so that $f_p(\theta^{1/p}) = \frac{2p}{\ln 1.152} (1.269)^{1/p} \le \frac{2p}{\ln 1.152} \cdot 1.269 \le 17.94 p < 18 p$. This proves the claim.
    \end{enumerate}
\end{proof}

We also remark that our reduction of $L_p$~TSP to Segmented-TSP can be generalized to multiple vehicles, with arbitrary start locations. Similar to the approach in \Cref{sec:complx} we can convert any optimal multi-vehicle solution to repetition of $O(\varepsilon^{-2})$ of prefix routes for each vehicle, all synchronized with a single $j \in \{0, \cdots, k-1\}$, picked uniformly at random. \Cref{lem:loss} can be subsequently adapted to allow limiting the search space to solutions that have all vehicles at starting locations simultaneously at all $3\lambda_i$, with negligible degrade of the optima. Finally, adapting the dynamic programming, we can guarantee a multiplicative $O(\varepsilon)$ loss given an (approximate) solver for multi-vehicle segmented TSP with constant $O(\varepsilon^{-2})$ deadlines, and the corresponding total number of destinations to be visited (by at least one vehicle) until up to each. This is indeed fruitful as results on segmented-TSP also generalize to multi-vehicle variant, e.g., in tree metric or Euclidean metric. Our algorithms can be similarly adapted in the case where release dates are added for the destinations.

\section{Concluding Remarks}

We studied combinatorial optimization problems whose objectives can be more appropriate, efficient, fair, and adjustable, depending on the enormous applications of optimal routing/scheduling. For TSP and TRP, the analyses of approximation algorithms as well as complexity results heavily rely on the linearity of the objective function. Hence, TFP and more generally $L_p$~TSP pose further challenging problems and require new techniques to be developed.
We developed multiple techniques aimed towards high precision and/or scalable approximation of the optimal route.

We provided a high precision polynomial time reduction of $L_p$~TSP to segmented-TSP with only a constant number of deadlines for visiting the required number of destinations.
Our reduction enables approximation schemes for $L_p$~TSP on Euclidean as well as weighted tree metrics; this is yet another motivation to further study the segmented-TSP problem.

Next, we investigated the case where we do not know what norm is the best to optimize, but want to be approximately optimal with respect to any.
In this thread, we developed an algorithm for All-Norm TSP on general metrics with approximation factor $8+\varepsilon$. We also provided a first inapproximability result for All-Norm TSP.

We showed the kernel of our combinatorial algorithm can be optimized for performance with respect to a target norm, ensuring $4.27$ and $8/(p \ln 4)^{1/p}$ approximations for TFP and $L_p$~TSP respectively.

Finally, with the combinatorial algorithms falling short of addressing the important multi-vehicle variants of our problems, we developed linear programming rounding techniques with first constant approximation bounds for multi-vehicle $L_p$~TSP and matching guarantees with the combinatorial approach for the single vehicle problems.

In the end, in addition to further improving the approximation bounds for the problems under study, we mention but a few of the potential directions to further expand this theory.

\begin{itemize}
    \item Can $L_p$~TSP be reduced to sub-exponentially many instances of TSP, with constant degrade to the approximation factor? We believe not. \Cref{obs:TSPreduction} is much better than trying $n!$ cases but still super-exponential. 
    \item $L_1$~TSP, i.e., TRP is harder than $L_\infty$~TSP, at least on trees. For what $p$ is the $L_p$~TSP problem the hardest?
    \item Bounding the integrality gap of the linear programming relaxations remains critical.
    \item While TRP is strongly NP-hard on weighted trees, its complexity is unknown for caterpillars \cite{S02}. Similarly complexity of TFP and $L_p$~TSP remain unresolved even on such special cases.
    \item While we theoretically claimed $p = 2$ to be ideal for the Traveling Firefighter Problem, this suggestion should be given further justification / investigation  in practice.
    \item Further applications of $L_p$~TSP, for instance in optimal containment of spread of pandemics \cite{hartke2004attempting, tennenholtz2020sequential}, are plausible.
    \item The impossibility bound for approximate All-Norm TSP and further NP-hardness of finding one can perhaps be improved.
\end{itemize}

\bibliographystyle{amsalpha}
\bibliography{refs}

\appendix

\section{All-Norm Inapproximability Example}\label{sec:inapprox-example}

Our example for $1.78$ inapproximability of All-Norm TSP has similar structure as the exponential sequence presented in \Cref{sec:all-norm}, and is computationally tuned and verified. This suggests the best (worst) example may have a different structure. We present the locations of the destinations in $\mathbb{R}^1$ to be visited by a traveler who is starting at $x = 200$. It was previously presented as Figure \ref{fig-example}.

\begin{align*}
    V = \{ 0, 200, 202, 204, 206, 208, 210, 212, 214, 216, 217, 218, 219, 220,\\ 221, 222, 223, 224, 225, 226, 228, 230, 232, 234, 236, 238, 240, 242, 244, 246,\\ 250, 254, 258, 262, 266, 270, 274, 278, 282, 286, 289, 292, 295, 298, 301, 304,\\ 307, 310, 313, 316, 316, 316, 316, 316, 316, 316, 316, 316, 316, 316, 322, 328,\\ 334, 340, 346, 352, 358, 364, 370, 376, 382, 388, 394, 400, 406, 412, 418, 424,\\ 430, 436, 446, 456, 466, 476, 486, 496, 506, 516, 526, 536, 540, 544, 548, 552,\\ 556, 560, 564, 568, 572, 576, 595, 614, 633, 652, 671, 690, 709, 728, 747, 766,\\ 775, 784, 793, 802, 811, 820, 829, 838, 847, 856, 888, 920, 952, 984, 1016, 1048,\\ 1080, 1112, 1144, 1176, 1199, 1222, 1245, 1268, 1291, 1314, 1337, 1360, 1383,\\ 1406, 1519, 1632, 1745, 1858, 1971, 2084, 2197, 2310, 2423, 2536 \}
\end{align*}

\end{document}